\documentclass[twocolumn,superscriptaddress,amsmath,amssymb,aps,pra]{revtex4}
\usepackage{amsmath,amsthm,amssymb}
\usepackage{latexsym}
\usepackage{amscd,graphicx}
\usepackage[titletoc]{appendix}
\usepackage{epsfig}
\usepackage{epstopdf}
\usepackage[normalem]{ulem}
\usepackage[dvipsnames]{xcolor}
\usepackage[colorlinks=true,linkcolor=red,citecolor=green]{hyperref}

\newcommand{\abs}[1]{| #1 |}

\begin{document}
	\title{Optomechanical interface induced strong spin-magnon coupling}
	
	\author{Wei Xiong}
	\altaffiliation{xiongweiphys@wzu.edu.cn}
	\affiliation{Department of Physics, Wenzhou University, Zhejiang 325035, China}
	
	
	\author{Mingfeng Wang}
	\altaffiliation{mfwang@wzu.edu.cn}
	\affiliation{Department of Physics, Wenzhou University, Zhejiang 325035, China}

	\author{Guo-Qiang Zhang}
	\altaffiliation{zhangguoqiang@csrc.ac.cn}
	\affiliation{School of Physics, Hangzhou Normal University, Hangzhou 311121, China}
	
	\author{Jiaojiao Chen}
	\altaffiliation{jjchenphys@hotmail.com}
	\affiliation{School of Physics and Optoelectronics Engineering, Anhui University, Hefei 230601, China}
	
	
	\date{\today}
	
\begin{abstract}
	Strong long-distance spin-magnon coupling is essential for solid-state quantum information processing and single qubit manipulation. Here, we propose an approach to realize strong spin-magnon coupling in a hybrid optomechanical cavity-spin-magnon system, where the optomechanical system, consisting of two cavities coupled to a common high-frequency mechanical resonator, acts as quantum interface. By eliminating the mechanical mode, a position-position coupling and two-mode squeezing of two cavities are induced. In the squeezing representation, the spin-photon, magnon-photon and photon-photon coupling strengths are exponentially amplified, thus lower- and upper-branch polaritons (LBP and UBP) are generated by strongly coupled squeezed modes of two cavities. Utilizing the critical property of the LBP, the coupling between the spin qubit (magnon) and LBP is greatly enhanced, while the coupling between the spin qubit (magnon) and UBP is fully suppressed. In the dispersive regime, strong and tunable spin-magnon coupling is induced by the virtual LBP, allowing quantum state exchange between them. Our proposal provides a promising platform to construct magnon-based hybrid systems and realize solid-state quantum information processing with optomechanical interfaces.

\end{abstract}
	
	\maketitle

\section{Introduction}
Optomechanical systems (OMSs), emerged as potential candidate for quantum information science owing to numerous exciting prospects for fundamental research and applications~\cite{Metcalfe-2014}, have attracted great interest theoretically and experimentally~\cite{Aspelmeyer}. The simplest OMS is formed by a cavity with two mirrors and a mechanical resonator (MR). Different from traditional cavities with two fixed mirrors, the optomechanical cavity has one fixed mirror and the other movable mirror.  The radiation pressure proportional to the cavity photon number acting on the MR causes the movable mirror to vibrate. This vibration in turn changes the length of the cavity (and thus the frequency of the cavity modes) and gives rise to
a nonlinearly optomechanical coupling between the cavity and the mechanical modes. Such an unconventional interaction results in rich backaction effects including sensing~\cite{Schreppler-2014,Wu-2017,Gil-Santos-2020,Fischer-2019}, ground-state cooling~\cite{Chan-2011,Teufel-2011}, squeezed light generation~\cite{Purdy-2014,Safavi-Naeini-2013,Aggarwal-2020}, nonreciprocal transport~\cite{Xu-2019,Shen-2016}, optomechanically induced transparency~\cite{Kronwald-2013,Weis-2010,Liuy-2013}, coupling enhancement~\cite{Xiong-2021,Lu-2013,Xiong2-2021}, nonlinear behaviors ({e.g., bi- and tristability and chaos} )~\cite{Lu-2015,Xiong3-2016}, and high-order exceptional points~\cite{Xiong3-2021,Xiong4-202205}. These indicates OMSs can be regarded as interfaces for building hybrid quantum systems~\cite{Camerer-2011,Yin-2015,Rakhubovsky-2016,Tian-2013,Bemmett-2016,Stannigel-2011,Xuereb-2021,Pei-2021,Shandilya-2021} for processing more complicated quantum information tasks.

Solid-state spins such as nitrogen-vacancy centers in diamond~\cite{RS}, with high controllability~\cite{Doherty2013}, long-coherence time~\cite{Gill2013,Jelezko2004,Balasubramian2009}, stable ground states~\cite{Doherty2013}, are considered as one of promising platforms for investigating quantum information science and quantum computation~\cite{xiang2013,kurizki2015}. In general, these spins are weakly coupled to other quantum systems or surrounding environment due to small magnetic dipole~\cite{Xiong-2021,kubo2010,marcos2010,zhu2011,Twamley2010}. Fortunately, growing magnons in yttrium-iron-garnet (YIG) spheres~\cite{Rameshti-2021,Yuan-2021,Quirion-2019,Wang-2020,Awschalom-2021,kusminskiy-2016,haigh-2015,huebl-2013,tabuchi-2015,quirion-2020,zhang-2016,hei-2021,zhang-2019,wang-2016,wang-2018,Wangy-2022,zhanggq-2021,zhanggq-2019,Shen-2021,Jing-2021,Zhang-2022}, with small volume but high spin density, can strongly interact with spin qubits by reducing the size of the sphere from~$\sim$mm scale to $\sim$nm scale~\cite{neuman-2020,neuman1-2021,neuman2-2021}, leading to the birth of spin-magnon hybrid systems~\cite{neuman-2020,neuman1-2021,neuman2-2021,Skogvoll-2021,Xiong4-2022}. However, spins are required to be placed near the surface of the nanosphere, which imposes difficulty on manipulating a single spin qubit via external magnetic fields. Therefore, realizing long-range spin-magnon coupling becomes open questions~\cite{Skogvoll-2021,Xiong4-2022,trifunovic-2013,Fukami-2021}.

Motivated by this, we thoretically propose a hybrid OMS consisting of two driven cavities (labelled as cavity $a$ and $c$) coupled to a common high-frequency MR plus a single spin qubit and a YIG nanosphere hosting magnons, where the spin qubit is weakly coupled to the cavity $a$. Magnons in the YIG nanosphere are also assumed to be weakly coupled to photons in the cavity $c$. This is reasonable because the magnon-photon coupling decreases with reducing the size of the YIG sphere although strong coupling between magnons in {\it millimeter} sphere and photons in a cavity has been achieved experimentally~\cite{Tabuchi-2014,ZhangX-2014,Goryachev-2014,Bai-2015,Zhangd-2015}. By linearizing the optomechanical Hamiltonian with strong driving fields in the blue-detuned regime and eliminating the mechanical mode, a position-position coupling between two cavities is obtained. Simultaneously, two-photon effects for two cavities are introduced, which can lead to photon squeezing. In the squeezing frame, the spin-cavity, magnon-cavity, and cavity-cavity coupling strengths are exponentially enhanced. The enhanced cavity-cavity coupling generates two polaritons, i.e., the LBP  and UBP. When the cavity-cavity coupling approaches to a critical value, the LBP exhibits a critical behavior, namely, the frequency of the LBP is real (imaginary) when the cavity-cavity coupling is smaller (larger) than the critical coupling.  At this critical point (CP), the fields of two cavities can be approximately equivalent to the position operator of the LBP with a large zero-point fluctuation, thus the coupling between the LBP and both the spin qubit and magnons are greatly improved, while  the coupling between the UBP and both the spin qubit and magnons are fully suppressed. In the dispersive regime, the virtually excited LBP acts as interface to induce a strong long-range spin-magnon coupling with accessible parameters. This strong coupling allows quantum state exchange between the single spin qubit and magnon in the presence of decoherence. Our proposal provides a potential path to manipulate spin qubits and realize solid-state quantum information processing with a hybrid optomechanical interface.

This paper is organized as follows. In Sec.~\ref{sec2}, the
model is described and the corresponding Hamiltonian
is given. Then we study quantum criticality of two squeezed cavities in
Sec.~\ref{sec3}. In Sec.~\ref{sec4}, the effective Hamiltonian describes the coupling between the single spin qubit and magnon is given. Using accessible parameters, this coupling is estimated to be in the strong coupling regime, allowing quantum state exchange between the spin qubit and magnon in the presence of decoherence.
Finally, a conclusion is given in Sec.~\ref{sec5}.

\section{Model and Hamiltonian}\label{sec2}
\subsection{Model}
\begin{figure}
	\center
	\includegraphics[scale=0.38]{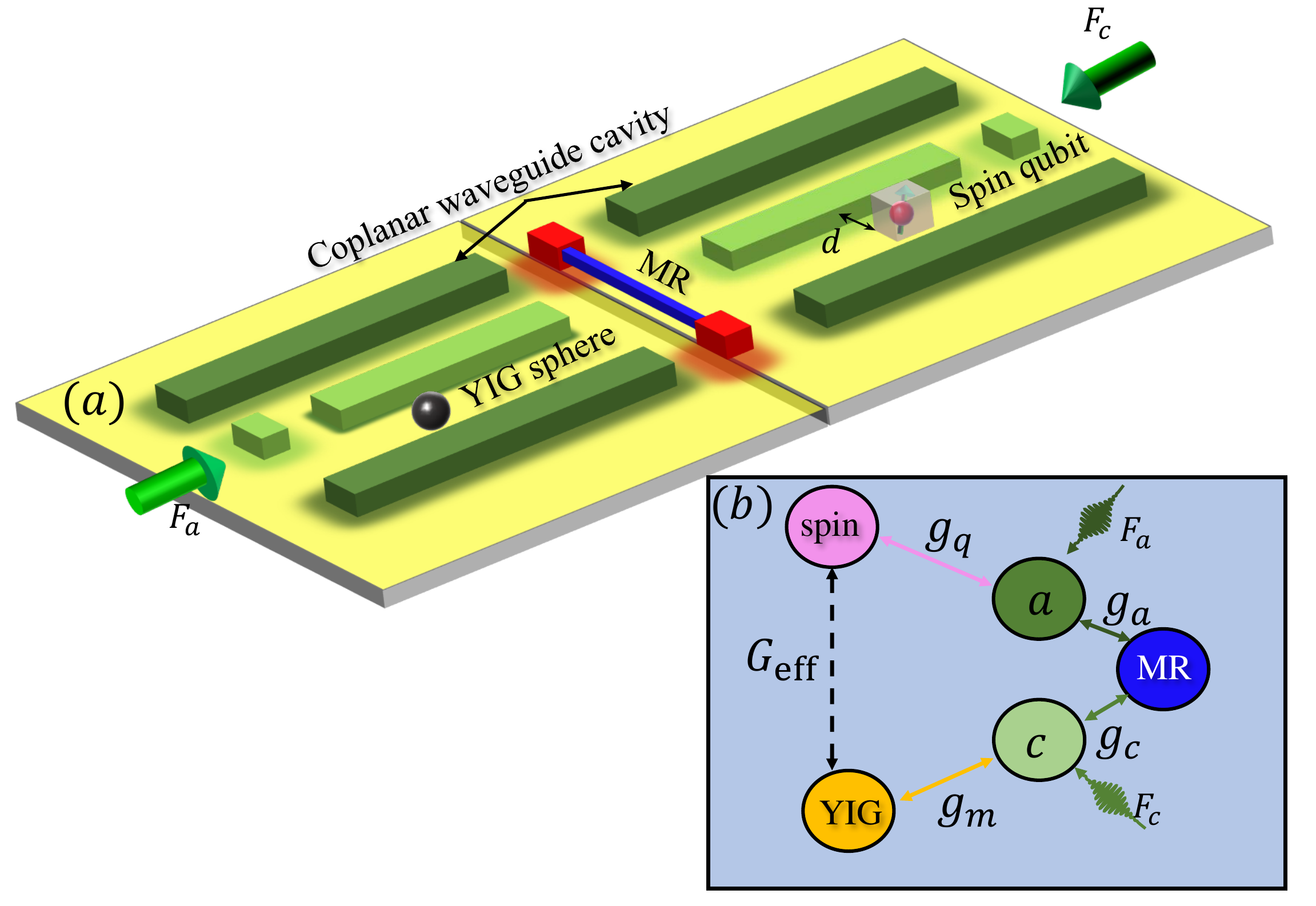}
	\caption{ (a) Schematic diagram of the proposed hybrid system consisting of two microwave cavities (e.g., superconducting coplanar waveguide resonator) coupled to a common high-frequency MR plus a single spin qubit and a YIG nanosohere hosting magnons, where two cavities are driven by two classical fields with amplitudes $F_a$ and $F_c$, and weakly interact with the spin qubit and magnon with coupling strengths $g_q$ and $g_m$, respectively. \{(b) Coupling diagram decribes the coupling between different subsystems. The solid lines denote the direct coupling, and the dashed line denotes the indirect coupling induced by the quantum interface}.\label{fig1}
\end{figure}
We consider a benchmark OMS consists of two microwave cavities (labelled as cavity $a$ and cavity $c$) coupled to a common MR with single-photon optomechanical coupling strengths $g_a$ and $g_c$ (see Fig.~\ref{fig1}). Such a setup has been demonstrated experimentally to achieve on-chip microwave circulators~\cite{Barzanjeh2017}. We here further consider the case that the cavity $a$ with eigenfrequency $\omega_a$ is weakly  coupled a single two-level spin qubit (e.g., nitrogen-vacancy center in diamond) with transition frequency $\omega_q$. Also, the cavity $c$ with eigenfrequency $\omega_c$ is weakly coupled to a magnon with frequency $\omega_m$ in a YIG nanosphere. Besides, two cavities are also externally driven by two classical fields with Rabi frequencies $F_a$ and $F_c$, respectively. In the rotating frame with respect to two classical fields, the total Hamiltonian of such a hybrid quantum system under the rotating-wave approximation can be written as {(see the details in the Appendix \ref{A})}
\begin{align}\label{eq1}
	H_{\rm total}=H_{\rm om}+H_q+H_m+H_d,
\end{align}
with
\begin{align}\label{eq2}
	H_{\rm om}=&\Delta_a a^\dag a+\omega_b b^\dag b+\Delta_c c^\dag c+(g_a a^\dag a-g_c c^\dag c)(b^\dag+b),\notag\\
	H_q=&\frac{1}{2}\Delta_q\sigma_z+g_q (a^\dag \sigma_-+a\sigma_+),\notag\\
	H_m=&\Delta_m m^\dag m+g_m(c^\dag m+m^\dag c),\notag\\
	H_d=&F_a (a^\dag+a) +F_c (c^\dag+c),
\end{align}
where $\Delta_{a(q)}=\omega_{a(q)}-\omega_a^d$ is the frequency detuning of the cavity $a$~(spin qubit) from the classical field with frequency $ \omega_a^d$ acting on the cavity $a$, $\Delta_{c(m)}=\omega_{c(m)}- \omega_c^d$ is the frequency detuning of the cavity $c$~(magnon) from the classical field with frequency $ \omega_c^d$ acting on the cavity $c$. {$g_q=2 g_e \mu_B B_{0,{\rm rms}}(d)$ \cite{Twamley2010} is the coupling strength between the spin qubit and the cavity $a$, where $B_{0,{\rm rms}}(d)=\mu_0 I_{\rm rms}/2\pi d$, with $\mu_0$ being the permeability of vacuum and $I_{\rm rms}=\sqrt{\hbar\omega_a/2L_a}$.
To estimate $g_q$, $\omega_a\sim 2\pi\times 2$ GHz and $L_a\sim 2$ nH are chosen \cite{Niemczyk2009}. For $d\sim50$~$\mu$m, $g_q\sim 2\pi\times 70$~Hz, and $d\sim50$~nm gives $g_q\sim 2\pi\times 7\sim20$~KHz. Obviously, the estimated spin-cavity coupling strength is smaller than the typical decay rate of the cavity with the gigahertz frequency and quality factor $Q\sim 3\times 10^4$ \cite{Niemczyk2009,Niemczyk2010}, i.e., $g_q<\kappa=\omega_a/Q\sim 1$ MHz. This indicates that the spin-cavity coupling is in the {\it weak-coupling} regime.} $g_m$ is the coupling strength between the magnon and the cavity $c$. Note that the strong coupling between the microwave cavity and the magnon in the millimeter YIG spheres has been achieved~\cite{Tabuchi-2014,ZhangX-2014,Goryachev-2014,Bai-2015,Zhangd-2015}. But the coupling between the cavity and the magnons in the nanometer YIG spheres have not been demonstrated, although the strong coupling between the sphere cavity and the nanomagnet has been theoretically studied. In fact, the magnon-cavity coupling decreases with reducing the size of the YIG sphere. For safety, we below assume the magnon-cavity coupling is also in the weak coupling regime.
\subsection{The effective total Hamiltonian}
Under the strong classical fields, the optomechanical Hamiltonian $H_{\rm om}$ in Eq.~(\ref{eq1}) can be linearized~\cite{Vitali2007}. {First, we rewrite} $a\rightarrow a+\langle a\rangle$ and $c\rightarrow c+\langle c\rangle$, where $\langle a\rangle~(\langle c\rangle)$ is the expectation value of the annihilation operator of the cavity $a$ ($c$) over its steady state, depending on Rabi frequencies of two calssical fields. {Then, we subsitute the above transformations into the dynamics of the optomechanical subsystem and neglect the higher-order fluctuation terms  (see the details in the Appendix \ref{A})}, the total Hamiltonian of the hybrid quantum system reduces to
\begin{align}\label{eq3}
	H_{L}=H_{\rm lin}+H_q+H_m,
	\end{align}
{where $H_q$ and $H_m$ are given by Eq.~({\ref{eq2}}), and
\begin{align}\label{eq4}
	H_{\rm lin}=&\Delta_a^\prime a^\dag a+\omega_b b^\dag b+\Delta_c^\prime c^\dag c\notag\\
	&+G_a(a^\dag+a)( b^\dag+b)+G_c(c^\dag+c)(b^\dag+b)
\end{align}
is the linearized optomechanical Hamiltonian.} Here $\Delta_{a(c)}^\prime=\Delta_{a(c)}\pm g_{a(c)}(\langle b\rangle+\langle b\rangle^*)$, with $\langle b\rangle$ the steady state value of the annihilation operator $b$, is the effective frequency detuning of the cavity $a~(c)$ induced by the displacement of the MR. $G_a=g_a \langle a\rangle$ and $G_c=-g_c \langle c\rangle$ are the enhanced optomechanical coupling strengths, which are assumed to be real for simplicity. This can be realized by tuning two classical fields. Note that the strong classical field $a$ ($c$) can indirectly give rise to the spin qubit flip (displacement of the magnon)~\cite{Xiong-2021}. Such effects can be offset by imposing the drive fields on the spin qubit (magnon).
\begin{figure}
	\center
	\includegraphics[scale=0.33]{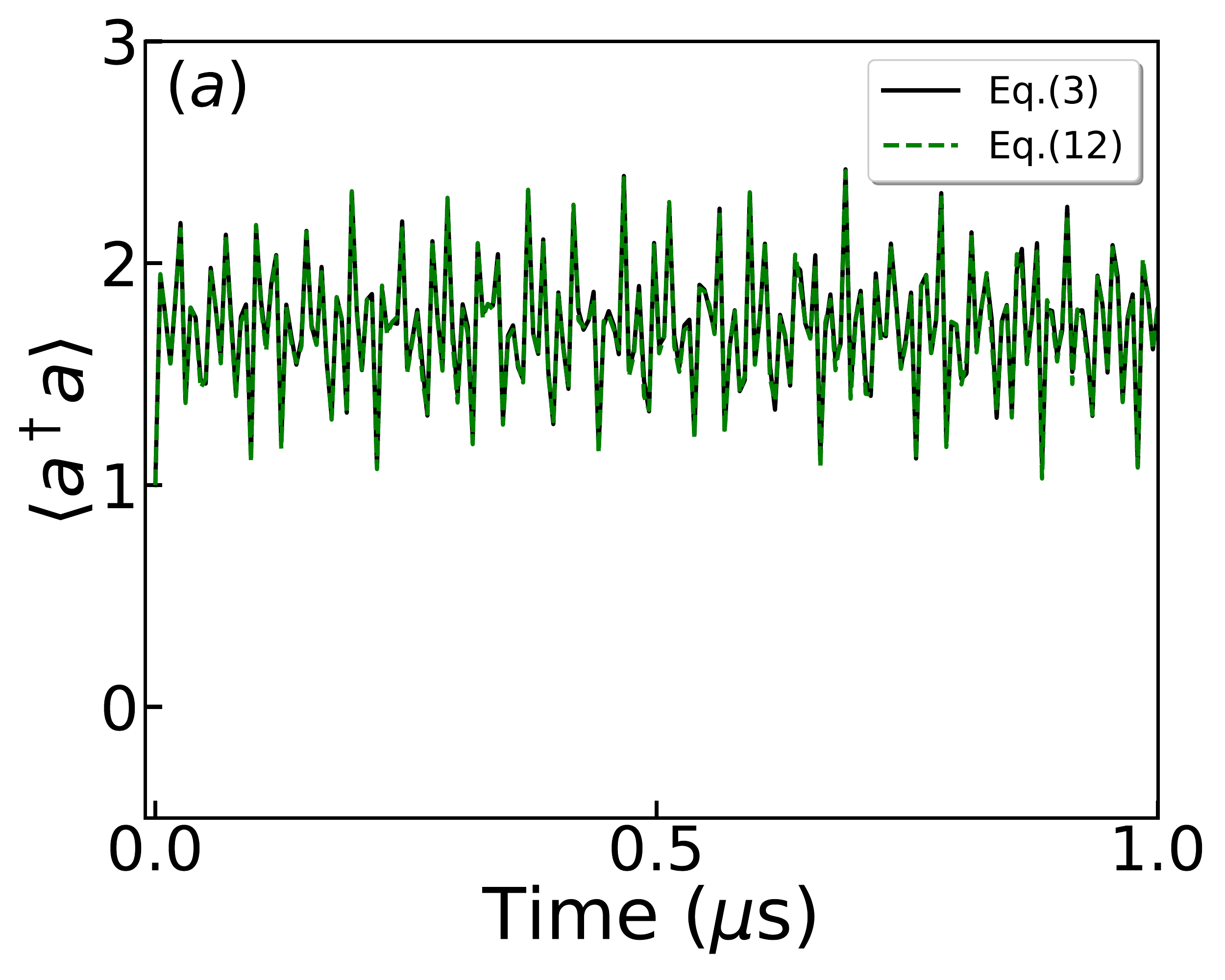}
	\includegraphics[scale=0.33]{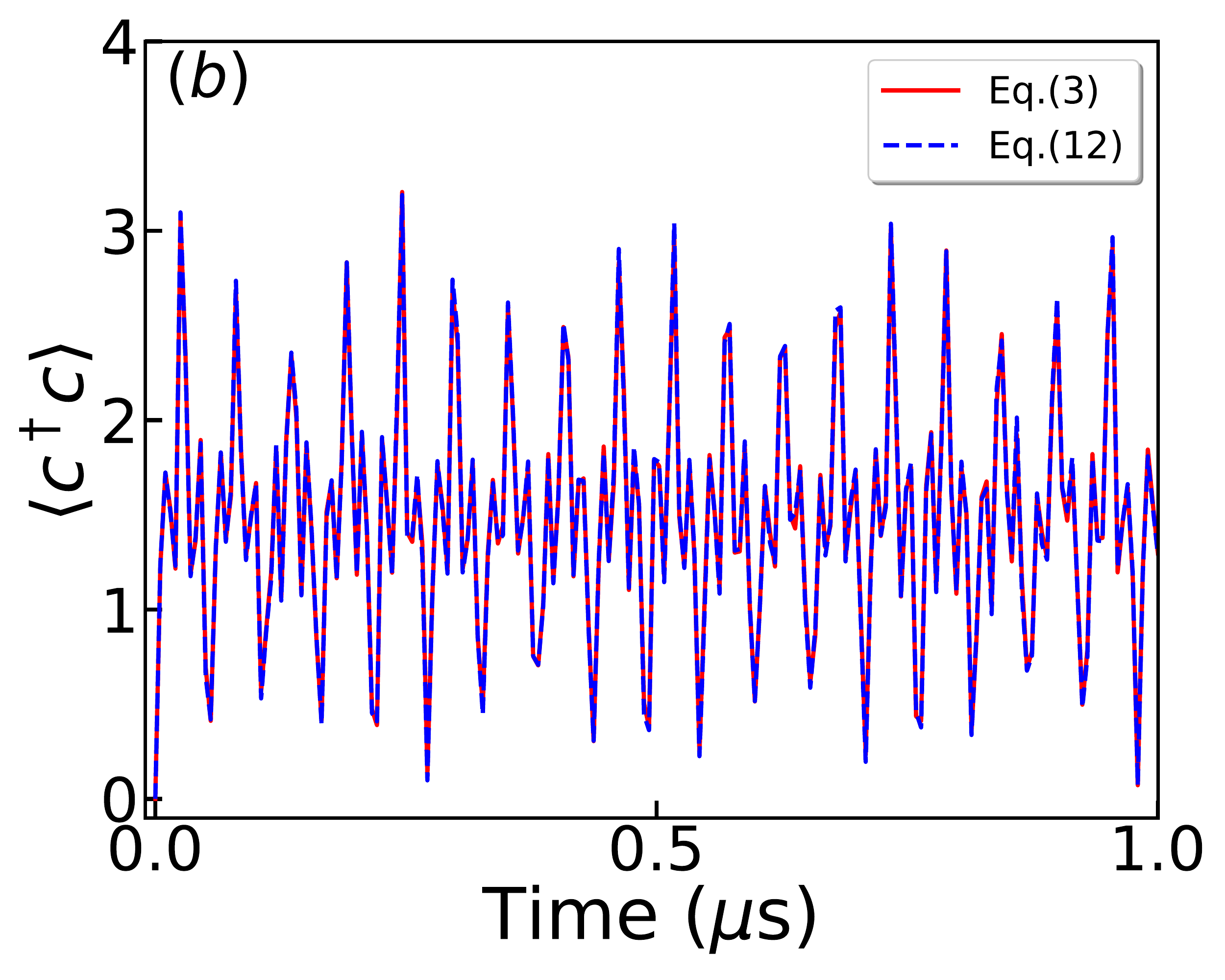}
	\caption{{(a) Simulating the expectation of photons in the cavity $a$ with Eqs.~(\ref{eq3}) and (\ref{eq8}). (b) Simulating the expectation of photons in the cavity $c$ with Eqs.~(\ref{eq3}) and (\ref{eq8}). The simulation results are respectively denoted by the solid and dashed curves in both (a) and (b).}}\label{figure2}
\end{figure}
{Below we consider that the MR is dispersively coupled to the two cavities, that is, the coupling strength is much smaller than the frequency detuning. For simplicity, we rewrite $H_{\rm lin}$ in Eq.~(\ref{eq4}) as
\begin{align}
	H_{\rm lin}=H_0+H_I,
\end{align}
where
\begin{align}
	H_0=&\Delta_a^\prime a^\dag a+\omega_b b^\dag b+\Delta_c^\prime c^\dag c,\\
	H_I=&G_a(a^\dag+a)( b^\dag+b)+G_c(c^\dag+c)(b^\dag+b).
\end{align}
We then apply the unitary transformation $U(\xi)=e^V$, with
\begin{align}
	V=&\xi_a^{(-)}(a^\dag b-a b^\dag)+\xi_a^{(+)}(a^\dag b^\dag-a b)\notag\\
	&+\xi_c^{(-)}(c^\dag b-c b^\dag)+\xi_c^{(+)}(c^\dag b^\dag-c b),
\end{align}
to the linearized Hamiltonian $H_L$ in Eq.~(\ref{eq3}), where $\xi_a^{(\pm)}=-G_a/\Delta_{\rm ab}^{(\pm)}$ with $\Delta_{\rm ab}^{(\pm)}=\Delta_a^\prime\pm\omega_b$, and $\xi_c^{(\pm)}=-G_c/\Delta_{\rm cb}^{(\pm)}$ with $\Delta_{\rm cb}{(\pm)}=\Delta_c^\prime\pm\omega_b$ are given by
\begin{align}
	[H_0,V]+H_I=0.
\end{align}
As $g_q$ and $g_m$ are the weak coupling strengths, so $H_{q}$ and $H_m$ are unchanged, while the linearized optomechanical Hamiltonian $H_{\rm lin}$ in Eq.~(\ref{eq3}) is transformed to~\cite{Frohlich,Nakajima}
\begin{align}
	H_{\rm ac}=&U^\dag H_{\rm lin} U	\approx H_0+\frac{1}{2}[H_I,V],
\end{align}
where higher-order terms are neglected and only the second-order terms are kept. This is reasonable because of $\xi_{a(c)}^{(\pm)}\ll1$ in the dispersive regime. By dropping the free energy of the mechanical mode,} $H_{\rm ac}$ can be specifically written as
\begin{align}\label{eqq11}
H_{\rm ac}=&\Delta_a^\prime a^\dag a+\Delta_c^\prime c^\dag c+G_{\rm ac}(a^\dag+a)(c^\dag+c)\notag\\
&+\frac{\chi_a}{2}(a^\dag+a)^2+\frac{\chi_c}{2}(c^\dag+c)^2,
\end{align}
where $G_{\rm ac}=\frac{1}{2}[G_a(\xi_c^{(+)}-\xi_c^{(-)})+G_c(\xi_a^{(+)}-\xi_a^{(-)})]$ is the indirect coupling strength induced by the mechanical mode, $\chi_{a(c)}=G_{a(c)}[\xi_{a(c)}^{(+)}-\xi_{a(c)}^{(-)}]$ is the coefficient of the mechanically induced second-order nonlinearity of the cavity $a$~($c$), which can squeezes the mode of the cavity $a$~($c$). Experimentally, room-temperature optomechanical squeezing using the blue-detuned driving has been achieved~\cite{Aggarwal2020}.

Thus, the total Hamiltonian in Eq.~(\ref{eq3}) can be effectively described as
\begin{align}\label{eq8}
	H_T=H_{\rm ac}+H_q+H_m,
\end{align}
{which is the approximate Hamiltonian of $H_L$ in Eq.~(\ref{eq3}). To show the resonability of our approximation in deducing the Hamiltonian $H_T$ in Eq.~(\ref{eq8}) from the Hamiltonian $H_L$ in Eq.~(\ref{eq3}), we simulate two Hamiltonians in Fig.~(\ref{figure2}) via studying the expectation of photons in two cavities, where we assume that initially the cavity $a$ is excited, the cavity $c$, the MR, the magnon and the spin qubit are all unexcited. The parameters are chosen as $\Delta_a^\prime=\Delta_c^\prime=-2\pi\times1$ kHz, $g_q=g_m=2\pi\times20$ kHz, $G_a=G_c=0.1\omega_b$ with $\omega_b=2\pi\times1$ GHz, $G_q=G_m=50$ MHz, and $\Delta_q=\Delta_m=10 G_q$. These parameters are also used to estimate spin-magnon coupling in Sec.~\ref{sec4}. From Fig.~\ref{figure2}, we can obviously find that our approximation is reasonable. The slight difference [see Fig.~\ref{figure2}(a) or \ref{figure2}(b)] between two Hamiltonians is due to the fact that higher-order terms in Eq.~(\ref{eq8}) are ignored.}

\section{Quantum criticality in the squeezing representation}\label{sec3}

{Note that the Hamiltonian $H_{\rm ac}$ in Eq.~(\ref{eq8}) or Eq.~(\ref{eqq11}) can squeeze photons in the cavities $a$ and $c$, so the the considered system can enter the squeezing representation by applying unitary transformations $U(r_a)=\exp[S(r_a)]$ with $S(r_a)=\frac{r_a}{2}(a^\dag a^\dag-a a)$ and $U(r_c)=\exp[S(r_c)]$ with $S(r_c)=\frac{r_c}{2}(c^\dag c^\dag-c c)$ to the Hamiltonian $H_T$ in Eq.~(\ref{eq8}), i.e.,
\begin{align}\label{e11}
	H_T^S=&U(r_c)^\dag U(r_a)^\dag H_T U(r_a)U(r_c)\notag\\
	=& H_{\rm ac}^S+H_q^S+H_m^S,
\end{align}
with
\begin{align}\label{eq11}
	H_q^S=&\frac{1}{2}\Delta_q\sigma_z+\frac{g_q}{2}e^{r_a}(a_s^\dag+a_s)(\sigma_++\sigma_-)\notag\\
	&-\frac{g_q}{2}e^{-r_a}(a_s^\dag-a_s)(\sigma_+-\sigma_-),\notag\\
	H_{m}^S=&\Delta_m m^\dag m+\frac{g_m}{2}e^{r_c}(c_s^\dag+c_s)(m+m^\dag)\\
	&-\frac{g_m}{2}e^{-r_c}(c_s^\dag-c_s)(m^\dag-m),\notag\\
	H_{\rm ac}^S=& \mathcal{W}_a a_s^\dag a_s+\mathcal{W}_c c_s^\dag c_s+\mathcal{G} (a_s^\dag+a_s)(c_s^\dag+c_s).\notag
\end{align}
Here $a_s=U(r_a) a U(r_a)^\dag=a\cosh(r_a)-a^\dag \sinh(r_a)$ and $c_s=U(r_c) c U(r_c)^\dag=c\cosh(r_c)-c^\dag \sinh(r_c)$ are used. Also, the squeezing parameters $r_a=\frac{1}{4}\ln({1+2\chi_a/\Delta_a^\prime})$ and  $r_c=\frac{1}{4}\ln({1+2\chi_c/\Delta_c^\prime})$ are chosen. In the Hamiltonian $H_{\rm ac}^S$, $\mathcal{W}_{a(c)}=\sqrt{\Delta_{a(c)}^\prime(\Delta_{a(c)}^\prime+\chi_{a(c)})}$ is the frequency of the squeezed photons in cavity $a~(c)$, and $\mathcal{G}=G_{\rm ac}e^{(r_a+r_c)}$ is the enhanced coupling strength between squeezed cavities $a$ and $c$ by the squeezing parameters $r_a$ and $r_c$}. The terms related to $e^{-r_{a(c)}}$ in Eq.~(\ref{eq11}) can be regarded as undesired corrections to the second terms in $H_{q(m)}^S$, which can be greatly suppressed when $e^{r_{a(c)}}\gg1$. To achieving $e^{r_{a(c)}}\gg1$, the following three conditions are required:
\begin{align}\label{c1}
	\omega_b&\gg\{\Delta_a^\prime,\Delta_c^\prime,G_a,G_c\},~\Delta_a^\prime,~\Delta_c^\prime<0,~
	\abs{\Delta_{a(c)}^\prime}\ll\abs{\chi_{a(c)}}.
\end{align}
Experimentally, high-frequency MRs have been prepared, so the first condition in Eq.~(\ref{c1}) can be satisfied. {In fact, the gigahertz MRs has been demonstrated for achieving microwave-optical photon conversion~\cite{Han2020,Zhao2022}} The second condition means that two cavities work {in the blue-detuned regime}, which can be realized by high-frequency classical fields. The third condition in Eq.~(\ref{c1}) can be achieved by increasing amplitudes of the driving fields. Thus, these three conditions can be accessed in our proposal.  {Under the conditions in Eq.~(\ref{c1}), $H_q^S$ and $H_m^S$ are respectively simplified as
\begin{align}\label{e15}
	\tilde{H}_q^S=&\frac{1}{2}\Delta_q\sigma_z+\frac{g_q}{2}e^{r_a}(a_s^\dag+a_s)(\sigma_++\sigma_-),\notag\\
	\tilde{H}_{m}^S=&\Delta_m m^\dag m+\frac{g_m}{2}e^{r_c}(c_s^\dag+c_s)(m+m^\dag).
\end{align}
When $g_q e^{r_a}/2\ll\Delta_q$ and $g_m e^{r_c}/2\ll\Delta_m$ are satisfied, the fast oscillating terms, such as $a_s^\dag\sigma_+$, $a_s\sigma_-$, $c_s^\dag m^\dag$ and $c_s m$, can be ignored via the rotating-wave approximation. So Eq.~(\ref{e15}) reduces to
\begin{align}\label{eqq16}
	\mathcal{H}_q^S=&\frac{1}{2}\Delta_q\sigma_z+\frac{g_q}{2}e^{r_a}(a_s^\dag\sigma_-+a_s\sigma_+)\notag\\
	\mathcal{H}_{m}^S=&\Delta_m m^\dag m+\frac{g_m}{2}e^{r_c}(c_s^\dag m+c_s m^\dag).
\end{align}
}We have also numerically verified the approximations leading from Eqs.~(\ref{e15}) to (\ref{eqq16}).
{For The Hamiltonian $H_{\rm ac}^S$ in Eq.~(\ref{eq11}), it is not difficult to find that $H_{\rm ac}^S$ can be fully diagnolized. The diagnolized Hamiltonian reads
\begin{align}\label{eqq18}
	H_{\rm AC}=\Omega_A A^\dag A+\Omega_C C^\dag C,
\end{align}
where
\begin{align}\label{eq16}
	A=&\frac{\cos\theta}{2\sqrt{\mathcal{W}_a\Omega_A}}(\Omega_A+\mathcal{W}_a)a_s+\frac{\cos\theta}{2\sqrt{\mathcal{W}_a\Omega_A}}(\Omega_A-\mathcal{W}_a)a_s^\dag\notag\\
	&-\frac{\sin\theta}{2\sqrt{\mathcal{W}_c\Omega_A}}(\Omega_A+\mathcal{W}_c)c_s-\frac{\sin\theta}{2\sqrt{\mathcal{W}_c\Omega_A}}(\Omega_A-\mathcal{W}_c)c_s^\dag,\notag\\
	C=&\frac{\sin\theta}{2\sqrt{\mathcal{W}_a\Omega_C}}(\Omega_C+\mathcal{W}_a)a_s+\frac{\sin\theta}{2\sqrt{\mathcal{W}_a\Omega_C}}(\Omega_C-\mathcal{W}_a)a_s^\dag\notag\\
	&+\frac{\cos\theta}{2\sqrt{\mathcal{W}_c\Omega_C}}(\Omega_C-\mathcal{W}_c)c_s+\frac{\cos\theta}{2\sqrt{\mathcal{W}_c\Omega_C}}(\Omega_C+\mathcal{W}_c)c_s^\dag,
\end{align}
with
\begin{align}
	\tan(2\theta)=\frac{4\mathcal{G}\sqrt{\mathcal{W}_a\mathcal{W}_c}}{\mathcal{W}_c^2-\mathcal{W}_a^2},
\end{align}
are the annihilation operators of the two new polaritons formed by the interaction between two squeezed cavities, respectively. $\Omega_A$ and $\Omega_C$, the eigenfrequencies of two new polaritons (i.e., the LBP and UBP) , are respectively given by
\begin{align}\label{eq15}
\Omega_A^2=&\frac{1}{2}\bigg[\mathcal{W}_a^2+\mathcal{W}_c^2-\sqrt{(\mathcal{W}_a^2-\mathcal{W}_c^2)^2+16\mathcal{G}^2\mathcal{W}_a\mathcal{W}_c}\bigg],\notag\\
\Omega_C^2=&\frac{1}{2}\bigg[\mathcal{W}_a^2+\mathcal{W}_c^2+\sqrt{(\mathcal{W}_a^2-\mathcal{W}_c^2)^2+16\mathcal{G}^2\mathcal{W}_a\mathcal{W}_c}\bigg].
\end{align}
Finally, the Hamiltonian $H_T^S$ in Eq.~(\ref{e11}) becomes
\begin{align}\label{eq22}
	\mathcal{H}_T^S=\mathcal{H}_q^S+\mathcal{H}_{m}^S+H_{\rm AC},
\end{align}
where $a_s$~($a_s^\dag$) and $c_s$~($c_s^\dag$) can be expressed in terms of $A,A^\dag,C$ and $C^\dag$ by solving Eq.~(\ref{eq16}).}
\begin{figure}
	\center
	\includegraphics[scale=0.33]{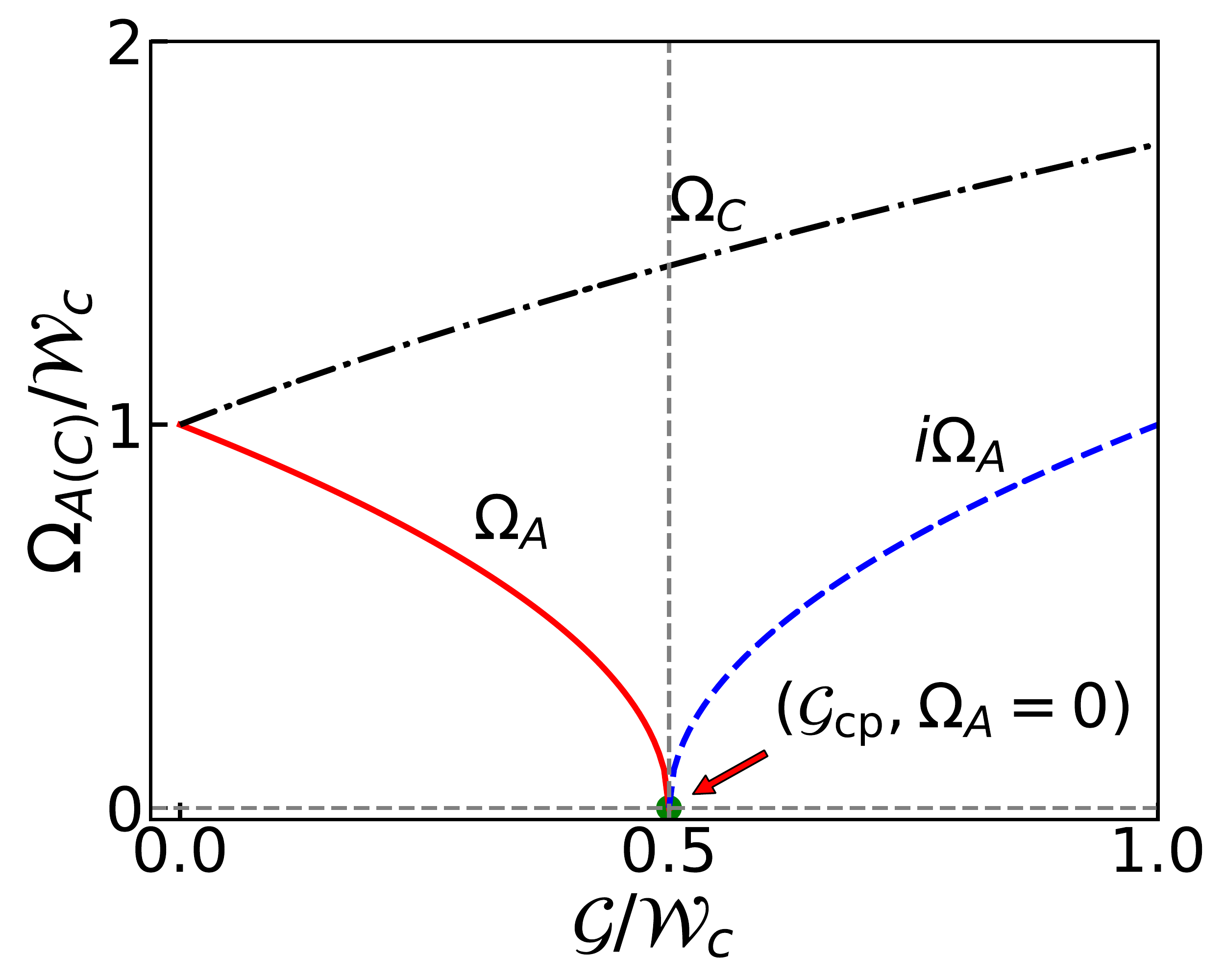}
	\caption{Eigenfrenquencies of the LBP  and UBP as functions of the coupling strength $\mathcal{G}$ between two squeezed cavities. Here the resonant case, $\mathcal{W}_a=\mathcal{W}_c$, is considered.}\label{fig2}
\end{figure}

To clearly show the criticality of the polaritons formed by two squeezed cavities,  we plot eigenfrencies $\Omega_A$ and $\Omega_C$ as functions of the coupling strength $\mathcal{G}$ between two squeezed cavities in Fig.~\ref{fig2}, where two squeezed cavities are resonant (i.e., $\mathcal{W}_a=\mathcal{W}_c$). Obviously, we can see that two polaritons have different behaviors by varying $\mathcal{G}$. Specifically, the UBP is always stable due to $\Omega_C^2>0$ for arbitrary $\mathcal{G}$ (see the black dashdotted curve), but the LBP is stable (i.e., $\Omega_A^2>0$) only when $\mathcal{G}<\mathcal{G}_{\rm cp}$ (see the red solid curve), where
\begin{align}\label{eq17}
	\mathcal{G}_{\rm cp}=\frac{1}{2}\sqrt{\mathcal{W}_a\mathcal{W}_c}.
\end{align}
For $\mathcal{G}>\mathcal{G}_{\rm cp}$, $\Omega_A^2<0$, corresponding to the unstable system (see the red dashed curve). Thus, the LBP has a critical behavior (i.e., $\Omega_A^2=0$) at the CP $\mathcal{G}=\mathcal{G}_{\rm cp}$~\cite{Xiong-2021,Lu-2013}, which can be realized in our considered hybrid system because $\mathcal{W}_a$, $\mathcal{W}_c$ and $\mathcal{G}$ are all controllable by tuning freqencies and amplitudes of classical fields. For the nonresonant case such as $\mathcal{W}_a=4\mathcal{W}_c>\mathcal{W}_c$, the CP is blueshifted at $\mathcal{G}_{\rm cp}=\mathcal{W}_c$. When $\mathcal{W}_a=0.01\mathcal{W}_c<\mathcal{W}_c$, the CP is redshifted at $\mathcal{G}_{\rm cp}=0.05\mathcal{W}_c$. {Note that when the decay rates from two squeezed modes are considered, not only the eigenvalues in Eq.~(\ref{eq15}) are changed,  but also the critical coupling strength in Eq.~(\ref{eq17}) is shifted. The specifically detailed discussions are given in the Appendix \ref{B}}.

\section{Strong spin-magnon coupling  mediated by the LBP }\label{sec4}

We now operate our hybrid system around the CP of two coupled squeezed-cavity modes by driving $\mathcal{G}$ approaching to $\mathcal{G}_{\rm cp}$, leading to $\Omega_A\rightarrow0$, $\Omega_C^2\rightarrow\mathcal{W}_a^2+\mathcal{W}_c^2$. We further consider the case that two squeezed cavities are resonant, i.e., $\mathcal{W}_a=\mathcal{W}_c\gg\Omega_A$, which gives $\mathcal{G}_{\rm cp}\rightarrow \frac{1}{2}\mathcal{W}_c$, $\Omega_C^2\rightarrow2\mathcal{W}_c^2$ and $\cos\theta=\sin\theta=1/\sqrt{2}$. Then the operators $a_s$ and $c_s$ around the CP are approximated as
\begin{align}\label{eq18}
	a_s\approx x_{\rm zpf} (A+A^\dag),~~
	c_s\approx-x_{\rm zpf} (A+A^\dag),
\end{align}
where $x_{\rm zpf}=\sqrt{{\mathcal{W}_c}/{(8\Omega_A)}}$, the zero-point fluctuation (ZPF) of the LBP, is greatly amplified by the extremely small $\Omega_A$ arising from the critical property of the LBP. {Substituting $a_s$ and $c_s$ given by Eq.~(\ref{eq18}) into the Hamiltonians $\mathcal{H}_q^S$ and $\mathcal{H}_m^S$ in Eq.~(\ref{eqq16}), we have
\begin{align}\label{eqq25}
	\mathcal{H}_q=&\frac{1}{2}\Delta_q\sigma_z+G_q(A+A^\dag)(\sigma_-+\sigma_+),\notag\\
	\mathcal{H}_m=&\Delta_m m^\dag m+G_m(A+A^\dag)(m+ m^\dag),
\end{align}}
\begin{figure}
	\center
	\includegraphics[scale=0.33]{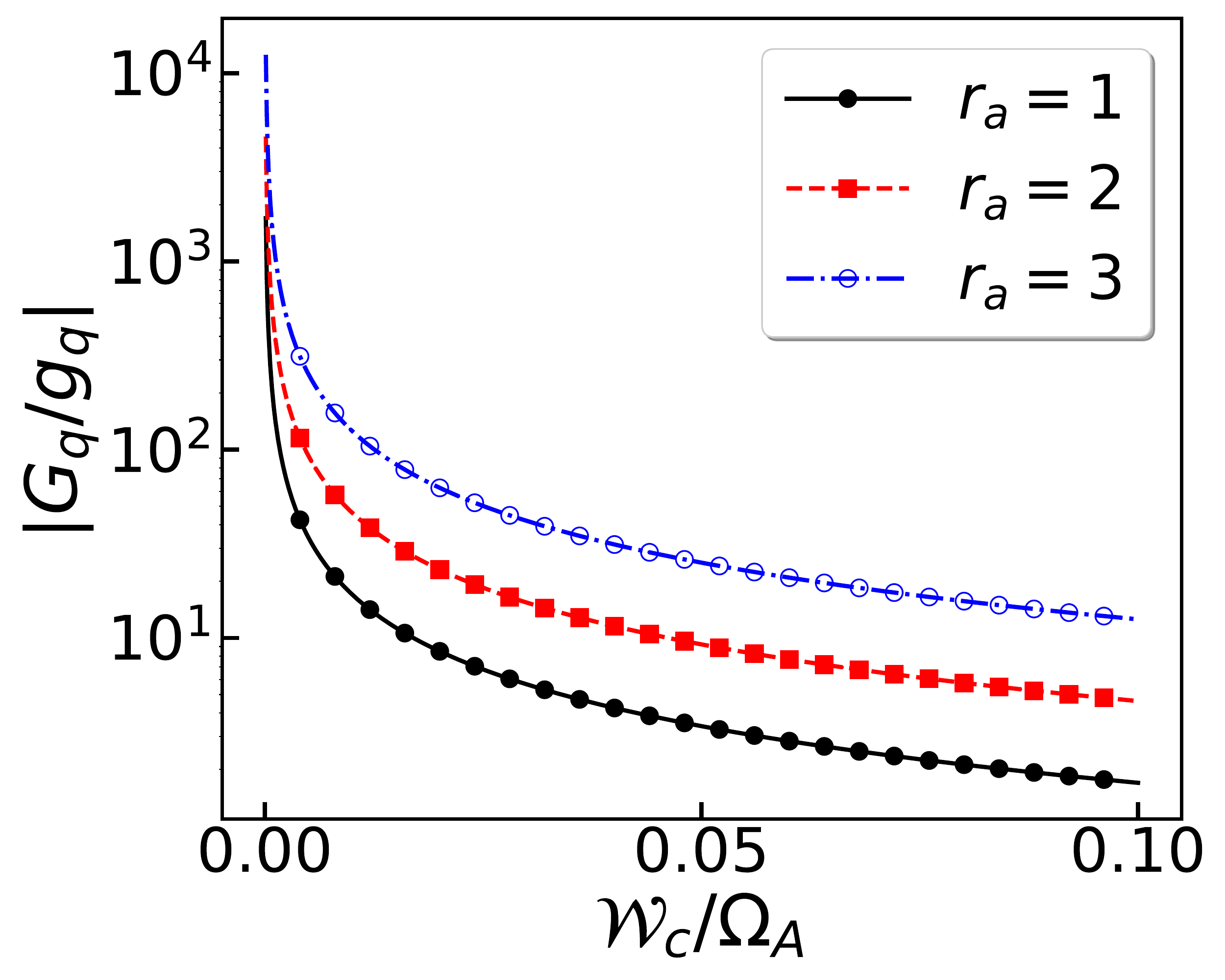}
	\caption{The relative coupling strength $\abs{G_q/g_q}$ vs the frequency of the LBP $\Omega_A$ in units of $\mathcal{W}_c$ with different squeezing parameters $r_a=1,2,3$.}\label{fig3}
\end{figure}
where $G_{q(m)}=\frac{1}{2} g_{q(m)} e^{r_{a(c)}} x_{\rm zpf}$ is the coupling strength between the spin qubit (magnon) and the LBP, as plotted in Fig.~\ref{fig3}. Around the CP, $x_{\rm zpf}$ is greatly enhanced due to the tiny $\Omega_A$, resulting in $G_{q}$ vastly improved (see the black curve). Moreover, $G_{q}$ can be further exponentially enhanced by increasing the squeezing parameter $r_a$ (see arbitrary two curves). From Fig.~\ref{fig3}, one can see that $G_{q}$ can be {\it three } or {\it four} orders of magnitude of the original coupling strength $g_q$ by reducing the frequency $\Omega_A$. In principle, $G_q$ can be arbitrary large by infinitely approaching $\Omega_A$ to zero. In the above analyses, the dissipations of the two squeezed cavities are not considered. When taking them into account, the coupling strength $G_{q(m)}$ can be still enhanced. In the presence of dissipations, the critical coupling strength $\mathcal{G}_{\rm cp}=\frac{1}{2}\sqrt{\mathcal{W}_a\mathcal{W}_c}$ is shifted to $\mathcal{G}_{\rm cp}^\prime=\frac{1}{2}\sqrt{(\mathcal{W}_a^2+\mathcal{K}^2)(\mathcal{W}_c^2+\mathcal{K}^2)/\mathcal{W}_a\mathcal{W}_c}$~\cite{Boneberg-2022}, where we assume that the decay rates of two squeezed cavities are equal, i.e., $\mathcal{K}_a=\mathcal{K}_b=\mathcal{K}$ (see the Appendix~\ref{B}). Around $\mathcal{G}=\mathcal{G}_{\rm cp}^\prime$, the coupling between the spin (magnon) and the LBP can be greatly enhanced due to the tiny $\Omega_A\rightarrow 0$. Thus, the strong spin-magnon coupling  can be obtained in the dispersive regime by adiabatically eliminating the degrees of freedom of the LBP~(see the Appendix~\ref{C}).{Combining Eqs.~(\ref{eqq25}) and (\ref{eqq18}), the total Hamiltonian in Eq.~(\ref{eq22}) within the rotating-wave approximation reduces to
\begin{align}\label{eq20}
	\mathcal{H}_T \equiv& \mathcal{H}_0+\mathcal{H}_I\notag\\
	=&\frac{1}{2}\Delta_q\sigma_z+\Delta_m m^\dag m+\Omega_A A^\dag A\notag\\
 	 &+G_q (A^\dag \sigma_-+A\sigma_+)+G_m (A^\dag m+A m^\dag),	 
\end{align}
where
\begin{align}
	\mathcal{H}_0=&\frac{1}{2}\Delta_q\sigma_z+\Delta_m m^\dag m+\Omega_A A^\dag A,\notag\\
	\mathcal{H}_I=&G_q (A^\dag \sigma_-+A\sigma_+)+G_m (A^\dag m+A m^\dag).
\end{align}
To ensure the validity of the rotating-wave approximation, $G_q/\Delta_q\ll1$ and $G_m/\Delta_m\ll1$ are required. As $\Omega_A$ is incredibly small, so we naturally have
\begin{align}\label{eq21}
	\zeta_q=\frac{G_q}{(\Delta_q-\Omega_A)}\ll1,~~~\zeta_m=\frac{G_m}{(\Delta_m-\Omega_A)}\ll1.
\end{align}
This indicates that the LBP is dispersively coupled to both the spin qubit and the magnons.  In this situation, the Fr\"{o}hlich-Nakajima transformation $\mathcal{U}=\exp(\mathcal{V})$ with
\begin{align}\label{eqq28}
	\mathcal{V}=\zeta_q (A^\dag \sigma_--A\sigma_+)+\zeta_m (A^\dag m-A m^\dag).
\end{align}
is allowed to deduce the indirect coupling between the spin qubit and the magnon via the LBP interface. Eq.~(\ref{eqq28}) directly give rise to
\begin{align}
	[\mathcal{H}_0,\mathcal{V}]+\mathcal{H}_I=0.
\end{align}
After the transformation and up to the second-order in $\zeta_q$ (or $\zeta_m$), the Hamiltonian $\mathcal{H}_T$ of the hybrid spin-magnon-polariton system in Eq.~(\ref{eq20}) reduces to
\begin{align}
	H_{\rm eff}^{\rm smp}=&\mathcal{U}^\dag \mathcal{H}_T\mathcal{U}\approx\mathcal{H}_0+\frac{1}{2}[\mathcal{H}_I,\mathcal{V}]\notag\\
	=&\frac{1}{2}\Delta_q\sigma_z+\Delta_m m^\dag m+\Omega_A A^\dag A\notag\\
	&+G_q\zeta_q(A^\dag A+\frac{1}{2})\sigma_z-G_m\zeta_m(A^\dag A-m^\dag m)\notag\\
	&+G_{\rm eff}(m^\dag \sigma_-+m\sigma_+),\label{eq23}
\end{align}}
where the term, i.e., $A^\dag A\sigma_z$, denotes the coupling between the spin qubit and the number of the LBP with the coupling strength $G_q\zeta_q$, the remain terms in the third line are the frequency shift caused by the transformation $\mathcal{U}$. The terms in the fourth line represent the indirect coupling between the magnon and the spin qubit with the {\it controllable} coupling strength $G_{\rm eff}=\frac{1}{2}(G_q\zeta_m+G_m\zeta_q)$. Due to the cross-Kerr-like coupling between the spin qubit and the LBP , the number of the LBP  will be unchanged, so we can eliminate the degrees of freedom of the LBP via replacing polariton number operator $A^\dag A$ by its expectation $\langle A^\dag A\rangle=N_A$, thus the spin-magnon-polariton Hamiltonian $H_{\rm eff}^{\rm smp}$ in Eq.~(\ref{eq23}) becomes
\begin{align}\label{eq24}
H_{\rm eff}=\frac{1}{2}\Delta_q^ {\rm eff}\sigma_z+\Delta_m^ {\rm eff} m^\dag m+G_{\rm eff}(m^\dag \sigma_-+m\sigma_+),
\end{align}
which is the effective spin-magnon Hamiltonian describing the couping between the spin qubit and the magnons. $\Delta_ q^ {\rm eff}=\Delta_q+G_q\zeta_q(2N_A+1)$ is the effective frequency of the spin qubit, induced by the polariton-dependent Stark shift $2N_AG_q\zeta_q$ and the zero-point energy $G_q\zeta_q$ of the LBP .  $\Delta_m^ {\rm eff}=\Delta_m+G_m\zeta_m$ is the effective frequency of the magnon induced by the dispersive coupling between the LBP  and magnon. At the single-polariton level, the Stark shift is $2G_q\zeta_q$, which can be observed due to the enhanced coupling strength $G_q$ around the CP (see Fig.~\ref{fig3}).
\begin{figure}
	\center
	\includegraphics[scale=0.33]{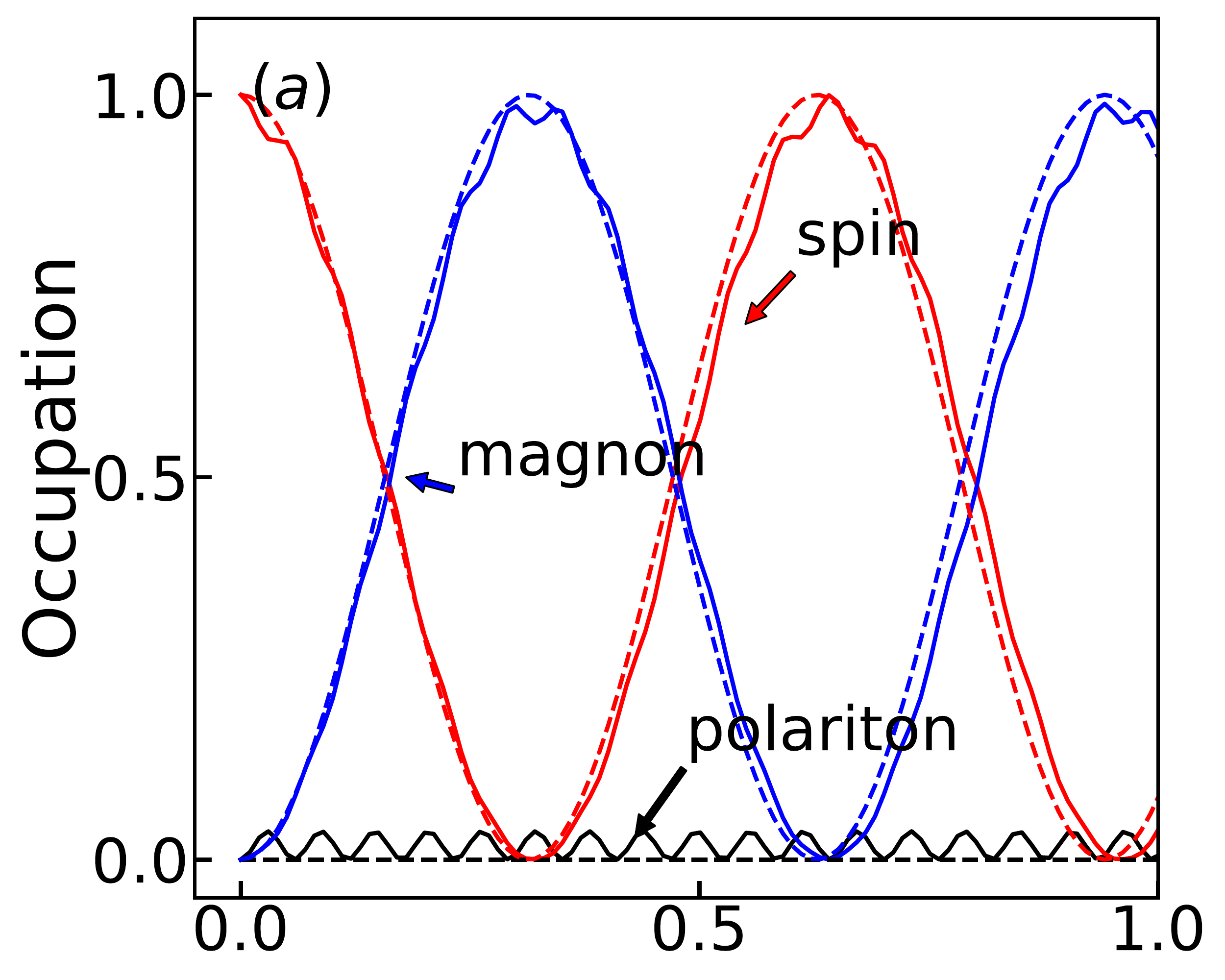}
	\includegraphics[scale=0.33]{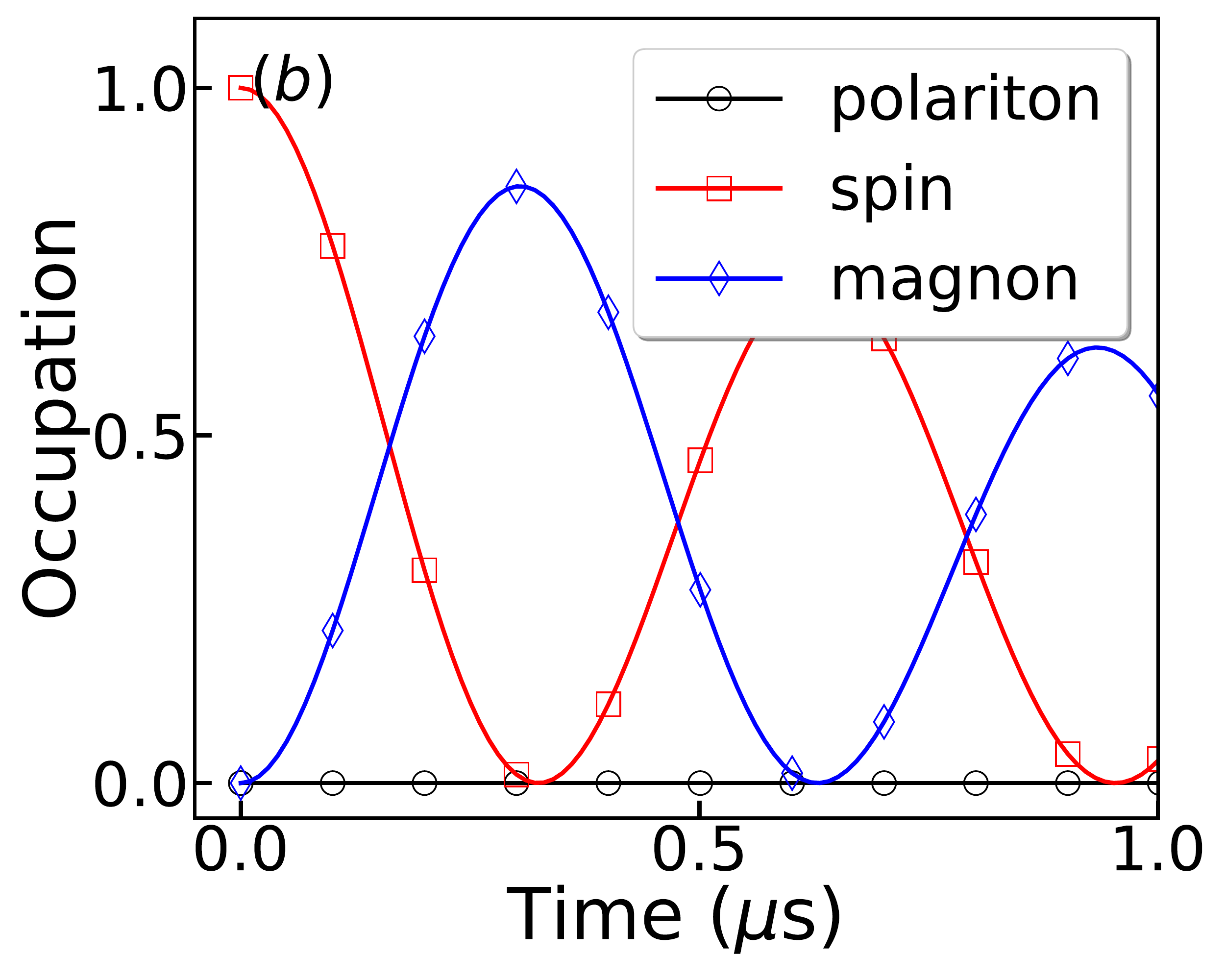}
	\caption{(a) Simulating the occupation of the LBP  and the spin qubit versus the evolution time with Eqs.~(\ref{eq20}) and (\ref{eq23}), respectively denoted by the solid and dashed curves. (b) The occupation of the LBP and the spin qubit versus the evolution time in the presence of dissipations from the magnon and qubits. In both (a) and (b), the polariton is in the ground state and the spin qubit is in the excited state.}\label{fig4}
\end{figure}

Below we give an estimation on the effective coupling strength $G_{\rm eff}$. For simplicity, we choose $\Delta_a^\prime=\Delta_c^\prime=-2\pi\times1$ kHz, $g_q=g_m=2\pi\times20$ kHz, and $G_a=G_c=0.1\omega_b$ with $\omega_b=2\pi\times1$ GHz, leading to $\chi_a=\chi_c=-125.7$ MHz. Thus, we have $r_a=r_c=2.65$ and $\mathcal{W}_a=\mathcal{W}_c=0.9$ MHz. These parameters ensures the conditions in Eq.~(\ref{c1}) is valid. By choosing $\mathcal{W}_c=10^6\Omega_A$, $x_{\rm zpf}=354$ is obtained, which gives rise to large $G_q=G_m=50$ MHz. To meet the dispersive conditons in Eq.~(\ref{eq21}), we take $\Delta_q=\Delta_m=10 G_q$, resulting in $G_{\rm eff}=5$ MHz, and $\Delta_q^{\rm eff}=\Delta_m^{\rm eff}\approx515$ MHz with $N_A=1$. For typical decay rate of the magnon in the YIG sphere is $\kappa_m\sim1$ MHz~\cite{Ballestero-2020}, thus $G_{\rm eff}>\kappa_m$. The decay rate of the nitrogen vacancy center spin is $\sim 1$ KHz~\cite{LiB-2019}, which is much smaller than $\kappa_m$. This indicates that the effective spin-magnon coupling enters {\it the strong coupling regime}, which allows to exchange quantum information between the spin qubit and magnon. With the above parametes, the polariton-dependent Stark shift is estimated as $\sim 10$ MHz at the single-polariton level, which is much larger than the decay rate of the magnon. Thus, it can be resolved in our proposal. {Also, the larger frequency of the low-frequency polariton is also allowed to realize strong spin-magnon coupling, such as $\Omega_A=10^{-4} \mathcal{W}_c\sim 0.1$~KHz , which leads to $G_q=G_m\sim17$ MHz. Then we tune $\Delta_a=\Delta_m=10G_{\rm q(m)}=170$ MHz, the effective spin-magnon coupling strength is $G_{\rm eff}\sim 1.7$ MHz, which is comparable to the typical decay rate of the magnon mode~\cite{Ballestero-2020}. In above discussion, we always assume that the magnon-cavity coupling is weak. But when a larger coupling strength of the cavity coupled to the YIG nanosphere is realized in future, the induced spin-magnon coupling will be much larger.}

The dynamics of the effective spin-magnon system  governed by Eq.~(\ref{eq24}) can be characterized by a master equation,
\begin{equation}
\frac{d\rho}{dt}=-i[H_{\rm eff},\rho]+\kappa_m \mathcal{D}[m]\rho
+\gamma_q\mathcal{D}[\sigma_-]\rho,\label{eq25}
\end{equation}
where $H_{\rm eff}$ is given by Eq.~(\ref{eq23}) or Eq.~(\ref{eq24}), $\mathcal{D}[o]\rho=o\rho o^\dag-\frac{1}{2}(o^\dag o\rho+\rho o^\dag o)$ for $o=m,\sigma_-$, and the longitudinal relaxation of the nitrogen vacancy spin qubit is neglected since it is much smaller than the transversal relaxation $\gamma_q$ in experiments~\cite{Angerer17}. Here the Hamiltonian in Eq.~(\ref{eq20}) is replaced by $H_{\rm eff}$ is reasonable because two Hamiltonians in the dispersive regime are approximately equivalent. This can be proven by the ideal dynamics of the spin qubit (the red curves), magnon (the blue curves), or LBP (the black curves) in Fig.~\ref{fig4}(a), where we assume that the spin qubit is in the excited state, and both the magnon and LBP are in their ground state. The small differences between Eqs.~(\ref{eq20}) and (\ref{eq23}), respectively denoted by the solid and dashed curves,  derive from the ignored higher-order terms in Eq.~(\ref{eq23}), which can be further remonved by choosing larger $\Delta_q$ and $\Delta_m$. From Fig.~\ref{fig4}(a), one can see that the LBP are nearly not excited, which indicates that the approximation of $A^\dag A=\langle A^\dag A\rangle$ in Eq.~(\ref{eq24}) is {suitable}.  In Fig.~\ref{fig4}(b), we plot the mean magnon number $\langle m^\dag m\rangle$ and the occupation probability of the spin qubit $\langle \sigma_z\rangle$ vs the evolution time, by numerically solving Eq.~(\ref{eq25}) with $\kappa_m=1$ MHz and $\gamma_q=1$ KHz, where the initial state is the same as in Fig.~\ref{fig4}(a). Obviously, the vacuum Rabi oscillation between the spin qubit and the magnon is observed, indicating that strong spin-magnon interaction is actually achieved in our proposal. With increasing the evolution time, this oscillation will be suppressed due to the dominant decay rate of the magnon.

\section{Conclusion}\label{sec5}

In summary, we have proposed an optomechanical interface for realizing a controllable and strong long-range coupling between a single spin qubit and a magnon in a YIG nanosphere, where the spin qubit and the magnon are both weakly coupled to two optomechanical cavities directly linked by a high-frequency MR. Eliminating the degrees of freedom of the MR, a strong position-position coupling and large second-order nonlinearities of two optomechanical cavities are obtained. These nonlinear effects give rise to cavity mode squeezing, thus the coupling strengths between the spin and one squeezed cavity, the magnon and the other squeezed cavity, and two squeezed cavities are exponentially enhanced. This strong photon-photon interaction generates two polaritons, i.e., the LBP  and UBP. By approaching photon-photon coupling strength to a critical value, the subsystem of the coupled cavities can exhibit critical behaviors predicted in the LBP. At this CP, the original operators of two cavities can be approximately expressed by the displacement operators of the LBP, thus the coupling between the spin (magnon) and the UBP can be fully suppressed, while the coupling between the spin (magnon) and the LBP  is greatly enhanced. In the dispersive regime, the LBP  can be adiabatically eliminated, leading to strong interaction between the spin and magnon estimated with accessible parameters. This strong coupling allows quantum state exchange between the spin and magnon, indicating that our proposal can provide a potential path to realize solid-state quantum information processing with optomechanical interface mediated spin-magnon systems.


This paper is supported by the key program of the Natural Science Foundation of Anhui (Grant No.~KJ2021A1301), and the National Natural Science Foundation of China (Grants No.~12205069 and No.~11804074). For numerical simulation, we used the QuTIP library~\cite{Johansson1,Johansson2}.

\setcounter{equation}{0}
\renewcommand{\theequation}{A\arabic{equation}}
\appendix{}
\section{The derivation of $H_{\rm lin}$ in Eq.~(\ref{eq4})}\label{A}
{In this appendix, we give a detailed derivation of the Hamitltonian $H_{\rm lin}$ in Eq.~(\ref{eq4}). The total Hamiltonian of the considered hybrid optomechanical system can be written as (setting $\hbar=1$)
\begin{align}\label{eqA1}
	H_{\rm total}^\prime=&\omega_a a^\dag a+\omega_b b^\dag b+\omega_c c^\dag c+(g_a a^\dag a-g_c c^\dag c)(b^\dag+b)\notag\\
	&+	\frac{1}{2}\omega_q\sigma_z+g_q (a^\dag \sigma_-+a\sigma_+)\notag\\
	&+\omega_m m^\dag m+g_m(c^\dag m+m^\dag c)\notag\\
	&+F_a (a^\dag e^{-i \omega_a^d t}+ae^{i \omega_a^d t})
	\notag\\
	& +F_c (c^\dag e^{-i \omega_c^d t}+ce^{i \omega_c^dt}),
\end{align}
where $\omega_{a(c)}$ is the frequency of the cavity $a$ ($c$) with the annihilation and creation operators $a$ ($c$) and $a^\dag~(c^\dag)$, $\omega_b$ is the frequency of the MR  with the annihilation (creation) operator $b$ ($b^\dag$), $\omega_q$ is the transition frequency of the spin qubit with the lowering (rising) operator $\sigma_-~(\sigma_+)$, and $\omega_m$ is the frequency of the magnon with the annihilation (creation) operator $m$ ($m^\dag$). The terms in the last two lines denote the interaction between the driving fields and the cavities $a$ and $(c)$, in which $\nu_{a(c)}$ is the frequency of the driving field $a~(c)$, and $F_{a(c)}$ is the corresponding amplitude. We further apply the unitary transformation $U=\exp[-i \omega_a^d(a^\dag a+\sigma_z)t-i \omega_c^d(c^\dag c+m^\dag m)t]$ to Eq.~(\ref{eqA1}), i.e.,
\begin{align}
	H_{\rm total}=&U^\dag H_{\rm total}^\prime U+i U \frac{\partial U^\dag}{\partial t}\notag\\
	=&H_{\rm om}+H_q+H_m+H_d,
\end{align}
which is just the main Hamiltonian given by Eq.~(\ref{eq1}) in the main text.}

Below we focus on the subsystem governed by the Hamiltonian $H_{\rm om}$.  The corresponding dynamics can be given by the quantum Langevin equation,
\begin{align}
	\partial a/\partial t=&-(\kappa_a+i\Delta_a)-ig_a a(b+b^\dag)+\sqrt{2\kappa_a}a_{\rm in},\notag\\
	\partial c/\partial t=&-(\kappa_c+i\Delta_c)+ig_c c(b+b^\dag)+\sqrt{2\kappa_c}c_{\rm in},\\
	\partial b/\partial t=&-(\kappa_b+i\omega_b)-ig_a a^\dag a+ig_c c^\dag c+\sqrt{2\kappa_b}b_{\rm in},\notag	
\end{align}
where $\kappa_o$ with $o=a,b,c$ is the decay rate, and $o_{\rm in}$ is the vacuum input noise. Under the strong driving condition, i.e., $\abs{\langle a\rangle},~\abs{\langle c\rangle}\gg1$, the above nonlinear equations can be linearized by writing $a\rightarrow a+\langle a\rangle$ and $c\rightarrow c+\langle c\rangle$. Neglecting the high-order fluctuation terms, the dynamics of the fluctuation operators reduces to
\begin{align}\label{eqA4}
	\partial a/\partial t=&-(\kappa_a+i\Delta_a^\prime)-iG_a (b+b^\dag)+\sqrt{2\kappa_a}a_{\rm in},\notag\\
	\partial c/\partial t=&-(\kappa_c+i\Delta_c^\prime)-iG_c (b+b^\dag)+\sqrt{2\kappa_c}c_{\rm in},\notag\\
	\partial b/\partial t=&-(\kappa_b+i\omega_b)-iG_a (a^\dag+a)-iG_c( c^\dag+ c)\notag\\&+\sqrt{2\kappa_b}b_{\rm in},
\end{align}
where $\Delta_a^\prime$ and $\Delta_c^\prime$ are the effective frequency detuning defined in Eq.~(\ref{eq4}), $G_a$ and $G_c$ are enhanced linearized optomechanical coupling strengths. By rewriting Eq.~(\ref{eqA4}) as $\partial o/\partial t=-i[o,H_{\rm lin}]-\kappa_o o+\sqrt{2\kappa_o}o_{\rm in}$, we have
{\begin{align}
	H_{\rm lin}=&\Delta_a^\prime a^\dag a+\omega_b b^\dag b+\Delta_c^\prime c^\dag c\notag\\
	&+G_a(a^\dag+a)( b^\dag+b)+G_c(c^\dag+c)(b^\dag+b),
\end{align}
which is just the linearized optomechanical Hamiltonian given by Eq.~(\ref{eq4}) in the main text.}

\renewcommand{\theequation}{B\arabic{equation}}

\section{The effect of the decay rate on the critical condition in Eq.~(\ref{eq17})}\label{B}
In the main text, the effects of the decay rates of the system on the critical point are not considered. Here,  we focus on these effects by investigating the dynamics of the subsystem Hamiltonian $H_{\rm ac}^S$, which can be given by the quantum Langevin equation,
\begin{align}
	\frac{\partial}{\partial t}R(t)=&-i[R,H_{\rm ac}^S]-\mathcal{K}_{R} R(t)-\sqrt{2\mathcal{K}_{R}}R_{\rm in}(t)\notag\\
	=&DR(t)-\sqrt{2\mathcal{K}_{R}}R_{\rm in}(t),
\end{align}
where $R(t)=(a_s,a^\dag_s,c_s,c_s^\dag)^T$,  $\mathcal{K}_R={\rm diag}(\mathcal{K}_a,\mathcal{K}_a,\mathcal{K}_c,\mathcal{K}_c)$ denotes the decay rates of the squeezed two cavities, $R_{\rm in}(t)=(a_{\rm s,in},a_{\rm s,in}^\dag,c_{\rm s,in},c_{\rm s,in}^\dag)^T$ are the Langevin forces, and the cofficient matrix
\begin{equation}\label{A2}
	D=\left(\begin{array}{cccc}
		-\mathcal{K}_a-i\mathcal{W}_a&0&-i\mathcal{G}&-i\mathcal{G}\\
		0&-\mathcal{K}_a+i\mathcal{W}_a&i\mathcal{G}&i\mathcal{G}\\
		-i\mathcal{G}&-i\mathcal{G}&-\mathcal{K}_c-i\mathcal{W}_c&0\\
	    i\mathcal{G}&i\mathcal{G}&0&-\mathcal{K}_c+i\mathcal{W}_c
	\end{array}\right).
\end{equation}
The corresponding cofficient matrix of the Hamiltonian $H_{\rm ac}^S$ is
\begin{align}
	\mathcal{D}=iD
\end{align}
The real and imaginary parts of the eigenvalues of $\mathcal{D}$ respectively represent the eigenfrequencies and the effective decay rates of the polariton modes. By including the decay rates, it is difficult to obtain the analytical expression of the eigenvalues of $\mathcal{D}$. But here we are only interested in the the effects of the decay rates on the critical condition in Eq.~(\ref{eq17}), so we assume $\mathcal{K}_{a}=\mathcal{K}_{c}=\mathcal{K}$ for analytically studying this effect. By solving the characteristic equation $\abs{\mathcal{D}-I*\Omega}=0$ with $\Omega$ the eigenvalues and $I$ the identity matrix, four specific eigenvalues are given by
\begin{align}\label{b4}
	\Omega_A^{(-)}=&{-i\mathcal{K}}-\sqrt{\frac{\mathcal{W}_a^2+\mathcal{W}_c^2}{2}-\sqrt{\frac{(\mathcal{W}_a^2-\mathcal{W}_c^2)^2+16\mathcal{G}^2\mathcal{W}_a\mathcal{W}_c}{4}}},\notag\\
	\Omega_A^{(+)}=&{-i\mathcal{K}}+\sqrt{\frac{\mathcal{W}_a^2+\mathcal{W}_c^2}{2}-\sqrt{\frac{(\mathcal{W}_a^2-\mathcal{W}_c^2)^2+16\mathcal{G}^2\mathcal{W}_a\mathcal{W}_c}{4}}},\notag\\
	\Omega_C^{(-)}=&{-i\mathcal{K}}-\sqrt{\frac{\mathcal{W}_a^2+\mathcal{W}_c^2}{2}+\sqrt{\frac{(\mathcal{W}_a^2-\mathcal{W}_c^2)^2+16\mathcal{G}^2\mathcal{W}_a\mathcal{W}_c}{4}}},\notag\\
	\Omega_C^{(+)}=&{-i\mathcal{K}}+\sqrt{\frac{\mathcal{W}_a^2+\mathcal{W}_c^2}{2}+\sqrt{\frac{(\mathcal{W}_a^2-\mathcal{W}_c^2)^2+16\mathcal{G}^2\mathcal{W}_a\mathcal{W}_c}{4}}}.
\end{align}
{Obviously, $[\Omega_{A(C)}^{(-)}]^2=[\Omega_{A(C)}^{(+)}]^2=\Omega_{A(C)}^2$ when $\mathcal{K}=0$, leading to $\Omega_A^2=0$ at $\mathcal{G}_{\rm cp}=\frac{1}{2}\sqrt{\mathcal{W}_a\mathcal{W}_c}$, which is just the critical point in the main text without including dissipation. At the ideal critical point $\mathcal{G}_{\rm cp}=\frac{1}{2}\sqrt{\mathcal{W}_a\mathcal{W}_c}$, $\Omega_A^{(-)}=\Omega_A^{(+)}=-i\mathcal{K}$, $\Omega_C^{(-)}=-i\mathcal{K}-\sqrt{\mathcal{W}_a^2+\mathcal{W}_c^2}$, and $\Omega_C^{(+)}=-i\mathcal{K}+\sqrt{\mathcal{W}_a^2+\mathcal{W}_c^2}$ for $\mathcal{K}\neq0$. This indicates that the criticality~(i.e., $\Omega^2=0$) in our proposal disappears at $\mathcal{G}_{\rm cp}=\frac{1}{2}\sqrt{\mathcal{W}_a\mathcal{W}_c}$ when the dissipation is included. To revisit the criticality of the considered system in the presence of dissipations, we can let $\Omega_A^{(-)}=0$, which directly results in~\cite{Boneberg-2022}
\begin{align}
	\mathcal{G}_{\rm cp}^\prime=\frac{1}{2}\sqrt{(\mathcal{W}_a^2+\mathcal{K}^2)(\mathcal{W}_c^2+\mathcal{K}^2)/\mathcal{W}_a\mathcal{W}_c}>\mathcal{G}_{\rm cp}.\label{b5}
\end{align}
This means the critical point in the main text is modified by the decay rates of two squeezed cavities.}

\renewcommand{\theequation}{C\arabic{equation}}
\section{The effect of the decay rate on the effective spin-magnon coupling strength $G_{\rm eff}$}\label{C}

{To clearly show coupling enhancement at the new critical point induced by the decay rates of two squeezed cavities, we diagonalize the cofficient matrix $D$ following the transformations in Eq.~(\ref{eq16}). As the decay rates of two squeezed cavities are included, the transformations~(\ref{eq16}) are non-unitary via replacing $\mathcal{W}_a$ and $\mathcal{W}_c$ by ${\mathcal{\tilde W}}_a=\mathcal{W}_a-i\mathcal{K}$ and ${\mathcal{\tilde W}}_c=\mathcal{W}_c-i\mathcal{K}$, respectively, where $\mathcal{K}_a=\mathcal{K}_c=\mathcal{K}$ is assumed. To be consistent with the main text, the resonant condition $\mathcal{W}_a=\mathcal{W}_c=\mathcal{W}$ is taken, so ${\mathcal{\tilde W}}_a={\mathcal{\tilde W}}_c=\mathcal{\tilde W}=\mathcal{W}-i\mathcal{K}$. With these assumptions, $\sin\theta=\cos\theta=1/\sqrt{2}$ is required for diagonalization. By solving Eq.~(\ref{eq16}), the reversible transformations in the presence of the decay rates can be expressed as
\begin{align}
a_s=&{\abs{a_+}}e^{i\phi_{a,+}} A+{\abs{a_-}}e^{i\phi_{a,-}} A^{\dagger}\notag\\
&+{\abs{c_+}}e^{i\phi_{c,+}} C+{\abs{c_-}}e^{i\phi_{c,-}} C^{\dagger} \notag\\
c_s=&-{\abs{a_+}}e^{i\phi_{a,+}} A-{\abs{a_-}} e^{i\phi_{a,-}}A^{\dagger}\notag\\
&+{\abs{c_+}}e^{i\phi_{c,+}} C+{\abs{c_-}}e^{i\phi_{c,-}}C^{\dagger},
\end{align}
with
\begin{align}\label{c2}
	\abs{a_\pm}=&\frac{\abs{\mathcal{\tilde W}}^2+|\Omega_A|^2\pm2(\mathcal{ K}{\rm Im[\Omega_A]}-\mathcal{ W}{\rm Re[\Omega_A]})}{2\abs{\mathcal{\tilde W}}|\Omega_A|},\notag\\
	\abs{c_\pm}=&\frac{\abs{\mathcal{\tilde W}}^2+|\Omega_C|^2\pm2(\mathcal{ K}{\rm Im[\Omega_C]}-\mathcal{ W}{\rm Re[\Omega_C]})}{2\abs{\mathcal{\tilde W}}|\Omega_C|},
\end{align}
where $a_\pm=\frac{{\mathcal{\tilde W}}\pm\Omega_A}{\sqrt{2{\mathcal{\tilde W}} \Omega_A}}=\abs{a_\pm}\exp(i\phi_{a,\pm})$ and $c_\pm=\frac{{\mathcal{\tilde W}}\pm\Omega_C}{\sqrt{2{\mathcal{\tilde W}} \Omega_C}}=\abs{c_\pm}\exp(i\phi_{c,\pm})$. Substituting $a_s$ and $c_s$ back into Eq.~(\ref{eq11}), the total Hamiltonian of the system becomes
\begin{align}
 \mathcal{H}=&\frac{1}{2}\Delta_q\sigma_z+\Delta_m m^\dag m+\Omega_A A^\dag A+\Omega_C C^\dag C\notag\\
 &+\left[(G_q^a e^{i\phi_{a,+}} A+G_q^c e^{i\phi_{c,+}} C)\sigma_++{\rm H.C.}\right]\notag\\
 &+\left[(G_m^a e^{i\phi_{a,+}} A+G_m^ce^{i\phi_{c,+}} C)m^\dag+{\rm H.C.}\right],\label{C3}
\end{align}
where $G_q^a=g_qe^{r_a}\abs{a_+}$, $G_q^c=g_qe^{r_a}\abs{c_+}$, $G_m^a=-g_me^{r_c}\abs{a_+}$, and $G_m^c=-g_me^{r_c}\abs{c_+}$. Different from Eq.~(\ref{eq20}), here the UBP is involved in Eq.~(\ref{C3}). This is because the couplings between the UBP and both the spin and magnon vanish around the ideal critical point $\mathcal{G}_{\rm cp}$ when the deday rates are not included.
\begin{figure}
	\center
	\includegraphics[scale=0.45]{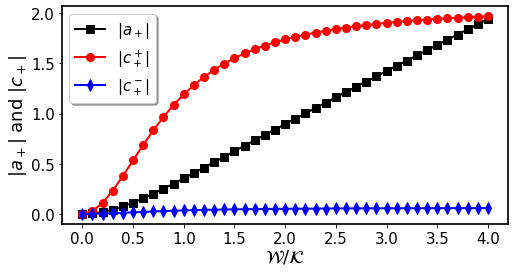}
	\caption{$\abs{a_+}$ and $\abs{c_+}$ in Eq.~(\ref{c4}) as functions of the ratio of $\mathcal{W}/\mathcal{K}$ at the ideal critical point $\mathcal{G}_{\rm cp}=\frac{1}{2}\mathcal{W}$ when the decay rates of two squeezed cavities are considered. The superscript '+~(-)' of $|c_+|$corresponds to the sign '+~(-)' of the second equation in Eq.~(\ref{c4}). Here $\mathcal{W}_{a}=\mathcal{W}_{c}$ and $\mathcal{K}_a=\mathcal{K}_a=\mathcal{K}$ are used.}\label{figs}
\end{figure} 
Below we give an estimation on $\abs{a_+}$ and $\abs{c_+}$. When the decay rates of two squeezed cavities are taken into account, $\Omega_A^{(-)}=\Omega_A^{(+)}=-i\mathcal{K}$, $\Omega_C^{(-)}=-i\mathcal{K}-\sqrt{2}\mathcal{W}$, and $\Omega_C^{(+)}=-i\mathcal{K}+\sqrt{2}\mathcal{W}$ are given at the ideal critical point $\mathcal{G}_{\rm cp}=\frac{1}{2}\sqrt{\mathcal{W}_a\mathcal{W}_c}=\frac{1}{2}\mathcal{W}$, it is easy to find that
\begin{align}\label{c4}
	\abs{a_+}=&\frac{\mathcal{W}^2}{2\mathcal{K}\sqrt{\mathcal{W}^2+\mathcal{K}^2}},\notag\\
	\abs{c_+}=&\frac{(3\pm2\sqrt{2})\mathcal{W}^2}{2\sqrt{(\mathcal{W}^2+\mathcal{K}^2)(\mathcal{K}+2\mathcal{W}^2)}}.
\end{align}
Obviously, both $a_+$ and $c_+$ are determinded by $\mathcal{K}$ when $\mathcal{W}$ is fixed. For microwave setup, $\mathcal{K}$ is general $\sim$ MHz, so $\mathcal{K}\sim\mathcal{W}$ in our proposal. Around $\mathcal{W/K}\sim1$, we can see that both $\abs{a_+}$ and $\abs{c_+}$ are small~(see curves in Fig.~\ref{figs}). This means the coupling between the polaritons and the spin (magnon) can not be significantly enhanced, finally giving rise to weak coupling between the spin and magnon. In a word, the proposal of coupling enhancement is fully washed out around the ideal critical point~$\mathcal{G}_{\rm cp}$ when the decay rates of two squeezed cavities are included. This is due to the fact that the critical point is shifted from $\mathcal{G}_{\rm cp}$~to $\mathcal{G}_{\rm cp}^\prime$ by the decay rates~[see Eq.~(\ref{b5})].

At $\mathcal{G}=\mathcal{G}_{\rm cp}^\prime$, $\Omega_A^{(-)}=0$, $\Omega_A^{(+)}=-2i\mathcal{K}$, $\Omega_C^{(-)}=-\sqrt{2\mathcal{W}^2+\mathcal{K}^2}-i\mathcal{K}$, and $\Omega_C^{(+)}=\sqrt{2\mathcal{W}^2+\mathcal{K}^2}-i\mathcal{K}$. According to the above discussions, we can see that $|a_+|$ and $\abs{c_+}$ can not be greatly enhanced for nonzero eigenvalues. But when  $\Omega_A^{(-)}\rightarrow0$, we can find that $|a_+|$ can be greatly strengthen due to the tiny $\Omega_A^{(-)}$, which is similar to the case of the coupling enhancement between the spin and magnon around the ideal critical point when the decay rates of two squeezed cavities are not taken into account. Therefore, $G_q^a$ and $G_m^a$ in Eq.~(\ref{C3}) can be significantly enhanced by $\abs{a_+}$, resulting in strong coupling between the spin and magnon via adiabatically eliminating the degrees of freedom of the polariton described by the operator $A$. Note that $G_q^c$ and $G_m^c$ can not be greatly enhanced, so the indirect coupling  between the spin and magnon can be ignored induced by the polariton described by the operator $C$.}


\begin{thebibliography}{99}
	
\bibitem{Metcalfe-2014} M. Metcalfe, Applications of cavity optomechanics, Appl. Phys. Rev. {\bf1}, 031105 (2014).

\bibitem{Aspelmeyer}M. Aspelmeyer, T. J. Kippenberg, and F. Marquardt, Cavity optomechanics, Rev. Mod. Phys. {\bf 86}, 1391 (2014).

\bibitem{Schreppler-2014}S. Schreppler, N. Spethmann, N. Brahms, T. Botter, M. Barrios, and D. M. Stamper-Kurn, Optically measuring force near the standard quantum limit. Science {\bf 344}, 1486 (2014).

\bibitem{Wu-2017}M. Wu, N. L.Y. Wu, T. Firdous, F. F. Sani, J. E. Losby, M. R. Freeman, and P. E. Barclay, Nanocavity optomechanical torque magnetometry and radiofrequency susceptometry, Nat. Nanotechnol. {\bf 12}, 127 (2017).

\bibitem{Gil-Santos-2020}E. Gil-Santos, J. J. Ruz, O. Malvar, I. Favero, A. Lema$\hat{\rm i}$tre, P. M. Kosaka, S. Garc$\acute{\rm i}$a-L$\acute{\rm o}$pez, M. Calleja, and J. Tamayo, Optomechanical detection of vibration modes of a single bacterium, Nat. Nanotechnol. {\bf 15}, 469 (2020).

\bibitem{Fischer-2019}R. Fischer, D. P. McNally, C. Reetz, G. G. T. Assumpç$\tilde{\rm a}$o, T. Knief, Y. Lin, and C. A. Regal, Spin detection with a micromechanical trampoline: Towards magnetic resonance microscopy harnessing cavity optomechanics, New J. Phys. {\bf 21}, 43049 (2019).

\bibitem{Chan-2011}J. Chan, T. P. M. Alegre, A. H. Safavi-Naeini, J. T. Hill, A. Krause, S. Gr\"{o}blacher, M. Aspelmeyer, and O. Painter, Laser cooling of a nanomechanical oscillator into its quantum ground state, Nature (London) {\bf 478}, 89 (2011).

\bibitem{Teufel-2011}J. D. Teufel, T. Donner, D. Li, J. W. Harlow, M. S. Allman, K. Cicak, A. J. Sirois, J. D. Whittaker, K. W. Lehnert, and R. W. Simmonds, Sideband cooling of micromechanical motion to the quantum ground state, Nature (London) {\bf 475}, 359 (2011).

\bibitem{Purdy-2014}T. P. Purdy, P. L. Yu, R. W. Peterson, N. S. Kampel, and C. A. Regal, Strong optomechanical squeezing of light, Phys. Rev. X {\bf 3}, 031012 (2013).

\bibitem{Safavi-Naeini-2013}A. H. Safavi-Naeini, S. Gr\"{o}blacher, J. T. Hill, J. Chan, M. Aspelmeyer, and O. Painter, Squeezed light from a silicon micromechanical resonator, Nature (London) {\bf 500}, 185 (2013).

\bibitem{Aggarwal-2020}N. Aggarwal, T. J. Cullen, J. Cripe, G. D. Cole, R. Lanza, A. Libson, D. Follman, P. Heu, T. Corbitt, and N. Mavalvala, Room-temperature optomechanical squeezing, Nat. Phys. {\bf 16}, 784 (2020).

\bibitem{Xu-2019}H. Xu, L. Jiang, A. A. Clerk, and J. G. E. Harris, Nonreciprocal control and cooling of phonon modes in an optomechanical system, Nature (London)  {\bf 568}, 65 (2019).

\bibitem{Shen-2016}Z. Shen, Y. L. Zhang, Y. Chen, C. L. Zou, Y. F. Xiao, X. B. Zou, F. W. Sun, G. C. Guo, and C. H. Dong, Experimental realization of optomechanically induced non-reciprocity, Nat. Photonics {\bf 10}, 657 (2016).

\bibitem{Kronwald-2013}A. Kronwald and F. Marquardt, Optomechanically Induced Transparency in the Nonlinear Quantum Regime, Phys. Rev. Lett. {\bf 111}, 133601 (2013).

\bibitem{Weis-2010} S. Weis, R. Riviere, S. Deleglise, E. Gavartin, O. Arcizet, A. Schliesser, and T. J. Kippenberg, Optomechanically Induced Transparency, Science {\bf 330}, 1520 (2010).

\bibitem{Liuy-2013}Y. Liu, M. Davanço, V. Aksyuk, and K. Srinivasan, Electromagnetically Induced Transparency and Wideband Wavelength Conversion in Silicon Nitride Microdisk Optomechanical Resonators, Phys. Rev. Lett. {\bf 110}, 223603 (2013).

\bibitem{Xiong-2021}W. Xiong, J. Chen, B. Fang, M. Wang, L. Ye, and J. Q. You, Strong tunable spin-spin interaction in a weakly coupled nitrogen vacancy spin-cavity electromechanical system, Phys. Rev. B {\bf 103}, 174106 (2021).

\bibitem{Lu-2013}X. Y. L$\ddot{\rm u}$, W. M. Zhang, S. Ashhab, Y. Wu, and F. Nori, Quantum-criticality-induced strong Kerr nonlinearities in optomechanical systems, Sci. Rep. {\bf 3}, 2943 (2013).

\bibitem{Xiong2-2021}J. Chen, Z. Li, X. Q. Luo, W. Xiong, M. Wang, and H. C. Li, Strong single-photon optomechanical coupling in a hybrid quantum system, Opt. Express {\bf 29}, 32639 (2021).

\bibitem{Xiong3-2016}W. Xiong, D. Y. Jin, Y. Qiu, C. H. Lam, and J. Q. You, Cross-Kerr effect on an optomechanical system, Phys. Rev. A {\bf 93}, 023844 (2016).

\bibitem{Lu-2015}X. Y. L\"{u}, H. Jing, J. Y. Ma, and Y. Wu, ${\mathcal PT}$-Symmetry Breaking Chaos in Optomechanics, Phys. Rev. Lett. {\bf 114}, 253601 (2015).

\bibitem{Xiong3-2021}W. Xiong, Z. Li, Y. Song, J. Chen, G. Q. Zhang, and M. Wang, Higher-order exceptional point in a pseudo-Hermitian cavity optomechanical system, Phys. Rev. A {\bf 104}, 063508 (2021).

\bibitem{Xiong4-202205}W. Xiong, Z. Li, G. Q. Zhang, M. Wang, H. C. Li, X. Q. Luo, and J. Chen, Higher-order exceptional point in a blue-detuned non-Hermitian cavity optomechanical system, arXiv:2205.07184.

\bibitem{Camerer-2011}S. Camerer, M. Korppi, A. J\"{o}ckel, D. Hunger, T. W. H\"{a}nsch, and P. Treutlein, Phys. Rev. Lett. {\bf 107}, 223001 (2011).

\bibitem{Yin-2015}Z. Yin, W. L. Yang, L. Sun, and L. M. Duan, Quantum network of superconducting qubits through an optomechanical interface, Phys. Rev. A {\bf 91}, 012333 (2015).

\bibitem{Rakhubovsky-2016}A. A. Rakhubovsky, N. Vostrosablin, and R. Filip, Squeezer-based pulsed optomechanical interface, Phys. Rev. A {\bf 93}, 033813 (2016).

\bibitem{Tian-2013}L. Tian, Robust Photon Entanglement via Quantum Interference in Optomechanical Interfaces, Phys. Rev. Lett. {\bf 110}, 233602 (2013).

\bibitem{Bemmett-2016}J. S. Bennett, K. Khosla, L. S. Madsen, M. R. Vanner, H. Rubinsztein-Dunlop, and W. P Bowen, A quantum optomechanical interface beyond the resolved sideband limit, New J. Phys. {\bf 18}, 053030 (2016).

\bibitem{Stannigel-2011}K. Stannigel, P. Rabl, A. S. S\o rensen, M. D. Lukin, and P. Zoller, Optomechanical transducers for quantum-information processing, Phys. Rev. A {\bf 84}, 042341 (2011).

\bibitem{Xuereb-2021}A. Xuereb, C. Genes, and A. Dantan, Strong Coupling and Long-Range Collective Interactions in Optomechanical Arrays, Phys. Rev. Lett. {\bf 109}, 223601 (2012).

\bibitem{Pei-2021}H. Xi and P. Pei, Quantum state transfer between distant optomechanical interfaces via shortcut to adiabaticity, Phys. Rev. A {\bf 104}, 052421 (2021).

\bibitem{Shandilya-2021}P. K. Shandilya, D. P. Lake, M. J. Mitchell, D. D. Sukachev, and  P. E. Barclay, Optomechanical interface between telecom photons and spin quantum memory, Nat. Phys. {\bf 17}, 1420 (2021).

\bibitem{RS}R. Schirhagl, K. Chang, M. Loretz, and C. L. Degen, Nitrogen-Vacancy Centers in Diamond: Nanoscale Sensors for Physics and Biology, Annu. Rev. Phys. Chem. {\bf 65}, 83 (2014).

\bibitem{Doherty2013}M. W. Doherty, N. B. Manson, P. Delaney, F. Jelezko, J. Wrachtrupe, L. C. L. Hollenberg, The nitrogen-vacancy colour centre in diamond, Phys. Rep. {\bf 528}, 1 (2013).

\bibitem{Gill2013}N. Bar-Gill, L. Pham, A. Jarmola, D. Budker, and R. Walsworth, Solid-state electronic spin coherence time approaching one second. Nat. Commun. {\bf4}, 1743 (2013).

\bibitem{Jelezko2004}F. Jelezko, T. Gaebel, I. Popa, A. Gruber, and J. Wrachtrup, Observation of Coherent Oscillations in A Single Electron Spin, Phys. Rev. Lett. {\bf92}, 076401 (2004).

\bibitem{Balasubramian2009}G. Balasubramian, P. Neumann, D. Twitchen, M. Markham, R. Koselov, N. Mizuochi, J. Isoya, J. Achard, J. Beck, J. Tissler, V. Jacques, P. R. Hemmer, F. Jelezko, and J. Wrachtrup, Ultralong spin coherence time in isotopically engineered diamond, Nat. Mater. {\bf8}, 383 (2009).

\bibitem{xiang2013}Z. L. Xiang, S. Ashhab, J. Q. You, and F. Nori, Hybrid quantum circuits: Superconducting circuits interacting withother quantum systems, Rev. Mod. Phys. {\bf85}, 623 (2013).

\bibitem{kurizki2015}G. Kurizki, P. Bertet, Y. Kubo, K. M\o{}lmer, D. Petrosyan, P. Rabl, and J. Schmiedmayer, Quantum technologies with hybrid systems, Proc. Natl. Acad. Sci. USA {\bf 112}, 3866 (2015).

\bibitem{kubo2010}Y. Kubo, F. R. Ong, P. Bertet, D. Vion, V. Jacques, D. Zheng, A. Dr$\rm \acute{e}$au, J.-F. Roch, A. Auffeves, F. Jelezko, J. Wrachtrup, M. F. Barthe, P. Bergonzo, and D. Esteve, Strong Coupling of A Spin Ensemble to A Superconducting Resonator, Phys. Rev. Lett. {\bf105}, 140502 (2010).

\bibitem{marcos2010}D. Marcos, M. Wubs, J. M. Taylor, R. Aguado, M. D. Lukin,and A. S. S\o rensen, Coupling Nitrogen-Vacancy Centers in Diamond to Superconducting Flux Qubits, Phys. Rev. Lett. {\bf105}, 210501 (2010).

\bibitem{zhu2011} X. Zhu, S. Saito, A. Kemp, K. Kakuyanagi, S. Karimoto, H. Nakano, W. J. Munro, Y. Tokura, M. S. Everitt, K. Nemoto, M. Kasu, N. Mizuochi, and K. Semba, Coherent coupling of a superconducting flux qubit to an electron spin ensemble in diamond, Nature (London) {\bf478}, 221 (2011).

\bibitem{Twamley2010}J. Twamley and S. D. Barrett, Superconducting cavity bus for single nitrogen-vacancy defect centers in diamond, Phys. Rev. B {\bf81}, 241202(R) (2010).

\bibitem{Rameshti-2021}B. Z. Rameshti, S. V. Kusminskiy, J. A. Haigh, K. Usami, D. Lachance-Quirion, Y. Nakamura, C. M. Hu, H. X. Tang, G. E. W. Bauer, and Y. M. Blanter, Cavity magnonics, arXiv: 2106.09312.

\bibitem{Yuan-2021} H. Y. Yuan, Y. Cao, A. Kamra, R. A. Duine, and P. Yan, Quantum magnonics: when magnon spintronics meets quantum information science, arXiv: 2111.14241.

\bibitem{Quirion-2019}D. Lachance-Quirion, Y. Tabuchi, A. Gloppe, K. Usami, and Y. Nakamura, Hybrid quantum systems based on magnonics, Appl. Phys. Express {\bf 12}, 070101 (2019).


\bibitem{Wang-2020}Y. P. Wang and C.-M. Hu, Dissipative couplings in cavity magnonics, J. Appl. Phys. {\bf 127}, 130901 (2020).

\bibitem{Awschalom-2021}D. D. Awschalom, C. H. R. Du, R. He, F. J. Heremans, A. Hoffmann, J. T. Hou, H. Kurebayashi, Y. Li, L. Liu, V. Novosad, J. Sklenar, S. E. Sullivan, D. Sun, H. Tang, V. Tiberkevich, C. Trevillian, A. W. Tsen, L. R. Weiss, W. Zhang, X. Zhang, L. Zhao, and  C. W. Zollitsch, Quantum engineering with hybrid magnonics systems and materials, IEEE Transactions on Quantum Engineering {\bf 2}, 5500836 (2021).

\bibitem{kusminskiy-2016}S. V. Kusminskiy, H. X. Tang and F. Marquardt, Coupled spin-light dynamics in cavity optomagnonics, Phys. Rev. A {\bf 94}, 033821 (2016).

\bibitem{haigh-2015}J. A. Haigh, S. Langenfeld, N. J. Lambert, J. J. Baumberg, A. J. Ramsay, A. Nunnenkamp, and A. J. Ferguson, Magneto-optical coupling in whispering-gallery mode resonators, Phys. Rev. A {\bf 92}, 063845 (2015).

\bibitem{huebl-2013}H. Huebl, C. W. Zollitsch, J. Lotze, F. Hocke, M. Greifenstein, A. Marx, R. Gross, and S. T. B. Goennenwein, High cooperativity in coupled microwave resonator ferrimagnetic insulator hybrids, Phys. Rev. Lett. {\bf 111}, 127003 (2013).

\bibitem{tabuchi-2015}Y. Tabuchi, S. Ishino, A. Noguchi, T. Ishikawa, R. Yamazaki, K. Usami, Y. Nakamura, Coherent coupling between a ferromagnetic magnon and a superconducting qubit, Science {\bf 349}, 6246 (2015).

\bibitem{quirion-2020}D. Lachance-Quirion,S. P. Wolski, Y. Tabuchi, S. Kono, K. Usami, and Y. Nakamura, Entanglement-based single-shot detection of a single magnon with a superconducting qubit, Science {\bf 367}, 425 (2020).

\bibitem{zhang-2016}X. Zhang, C. L. Zou, L. Jiang, and H. X. Tang, Cavity magnomechanics, Sci. Adv. {\bf 2}, e1501286 (2016).

\bibitem{hei-2021}X. L. Hei, X. L. Dong, J. Q. Chen, C. P. Shen, Y. F. Qiao, and P. B. Li, Enhancing spin-photon coupling with a micromagnet, Phys. Rev. A {\bf 103}, 043706 (2021).

\bibitem{zhang-2019}G. Q. Zhang, Y. P. Wang, and J. Q. You, Theory of the magnon
Kerr effect in cavity magnonics, Sci. China-Phys. Mech. Astron. {\bf 62}, 987511 (2019).

\bibitem{wang-2016}Y. P. Wang, G. Q. Zhang, D. Zhang, X. Q. Luo, W. Xiong, S.
P. Wang, T. F. Li, C. M. Hu, and J. Q. You, Magnon Kerr effect in a strongly coupled cavity-magnon system, Phys. Rev. B {\bf 94}, 224410 (2016).

\bibitem{wang-2018}Y. P. Wang, G. Q. Zhang, D. Zhang, T. F. Li, C. M. Hu, and J. Q. You, Bistability of Cavity Magnon-Polaritons, Phys. Rev. Lett. {\bf 120}, 057202 (2018).

\bibitem{zhanggq-2021}G. Q. Zhang, Z. Chen, W. Xiong, C. H. Lam, and J. Q. You, Parity-symmetry-breaking quantum phase transition in a cavity magnonic system driven by a parametric field, Phys. Rev. B {\bf 104}, 064423 (2021).

\bibitem{Wangy-2022}Y. Wang, W. Xiong, Z. Xu, G. Q. Zhang, and J. Q. You, Dissipation-induced nonreciprocal magnon blockade in a magnon-based hybrid system, Sci. China Phys. Mech. Astron. {\bf 65}, 1 (2022).

\bibitem{zhanggq-2019}G. Q. Zhang and J. Q. You, Higher-order exceptional point in a cavity magnonics system, Phys. Rev.  B {\bf 99}, 054404 (2019).

\bibitem{Shen-2021}R. C. Shen, Y. P. Wang, J. Li, S. Y. Zhu, G. S. Agarwal, and J. Q. You, Long-Time Memory and Ternary Logic Gate Using a Multistable Cavity Magnonic System, Phys. Rev. Lett. {\bf 127}, 183202 (2021).

\bibitem{Jing-2021}S. Qi and Jun Jing, Generation of Bell and GHZ states from a hybrid qubit-photon-magnon system, 	Phys. Rev. A {\bf 105}, 022624 (2022).

\bibitem{Zhang-2022}G. Q. Zhang, W. Feng, W. Xiong, Q. P. Su, and C. P. Yang, Generation of long-lived W states via reservoir engineering in dissipatively coupled systems, Phys. Rev. A {\bf 107}, 012410 (2023).


\bibitem{neuman-2020}T. Neuman, D. S. Wang, and P. Narang, Nanomagnonic Cavities for Strong Spin-Magnon Coupling and Magnon-Mediated Spin-Spin Interactions, Phys. Rev. Lett. {\bf 125}, 247702 (2020).

\bibitem{neuman1-2021} D. S. Wang, T. Neuman, and P. Narang, Spin Emitters beyond the Point Dipole Approximation in Nanomagnonic Cavities, J. Phys. Chem. C {\bf 125}, 6222 (2021).

\bibitem{neuman2-2021}D. S. Wang, M. Haas, and P. Narang, Quantum Interfaces to the Nanoscale, ACS Nano {\bf 15}, 7879 (2021).

\bibitem{Skogvoll-2021}I. C. Skogvoll, J. Lidal, J. Danon, and A. Kamra, Tunable anisotropic quantum Rabi model via magnon|spin-qubit ensemble, Phys. Rev. Applied {\bf 16}, 064008 (2021).

\bibitem{Xiong4-2022}W. Xiong, M. Tian, G. Q. Zhang, and J. Q. You, Strong long-range spin-spin coupling via a Kerr magnon interface, Physical Review B {\bf105}, 245310 (2022).

\bibitem{trifunovic-2013}L. Trifunovic, F. L. Pedrocchi, and D. Loss, Long-Distance Entanglement of Spin Qubits via Ferromagnet, Phys. Rev. X {\bf 3}, 041023 (2013).

\bibitem{Fukami-2021}M. Fukami, D. R. Candido, D. D. Awschalom, and M. E. Flatt$\acute{\rm  e}$, Opportunities for Long-Range Magnon-Mediated Entanglement of Spin Qubits via On- and Off-Resonant Coupling, PRX Quantum {\bf 2}, 040314 (2021).

\bibitem{Tabuchi-2014}Y. Tabuchi, S. Ishino, T. Ishikawa, R. Yamazaki, K. Usami, and Y. Nakamura, Hybridizing Ferromagnetic Magnons and Microwave Photons in the Quantum Limit, Phys. Rev. Lett. {\bf 113}, 083603 (2014).

\bibitem{ZhangX-2014}X. Zhang, C.-L. Zou, L. Jiang, and H. X. Tang, Strongly Coupled Magnons and Cavity Microwave Photons, Phys. Rev. Lett. {\bf 113}, 156401 (2014).

\bibitem{Goryachev-2014}M. Goryachev, W. G. Farr, D. L. Creedon, Y. Fan, M. Kostylev, and M. E. Tobar, High-Cooperativity Cavity QED with Magnons at Microwave Frequencies, Phys. Rev. Appl. {\bf 2}, 054002 (2014).

\bibitem{Bai-2015}L. Bai, M. Harder, Y. P. Chen, X. Fan, J. Q. Xiao, and C. M. Hu, Spin Pumping in Electrodynamically Coupled Magnon-Photon Systems, Phys. Rev. Lett. {\bf 114}, 227201 (2015).

\bibitem{Zhangd-2015} D. Zhang, X. M. Wang, T. F. Li, X. Q. Luo, W. Wu, F. Nori, and J. Q. You, Cavity quantum electrodynamics with ferromagnetic magnons in a small yttrium-iron-garnet sphere, npj Quantum Information {\bf 1}, 15014 (2015).

{\bibitem{Barzanjeh2017}S. Barzanjeh, M. Wulf, M. Peruzzo, M. Kalaee, P.B. Dieterle, O. Painter, and J.M. Fink, Mechanical on-chip microwave circulator,  Nat Commun {\bf8}, 953 (2017).}

{\bibitem{Niemczyk2009}T. Niemczyk, F. Deppe, M. Mariantoni, E. P. Menzel, E. Hoffmann, G. Wild, L. Eggenstein, A. Marx, and R. Gross, Fabrication technology of and symmetry breaking in superconducting quantum circuits, Supercond. Sci. Technol. {\bf22}, 034009 (2009)}.

{\bibitem{Niemczyk2010}T. Niemczyk, F. Deppe, H. Huebl, E. P. Menzel, F. Hocke, M. J. Schwarz, J. J. Garcia-Ripoll, D. Zueco, T. H$\rm \ddot{u}$mmer, E. Solano, A. Marx, and R. Gross, Circuit quantum electrodynamics in the ultrastrong-coupling regime, Nat. Phys. {\bf6}, 772 (2010).}

\bibitem{Vitali2007}D. Vitali, S. Gigan, A. Ferreira, H. R. B\"{o}hm, P. Tombesi, A. Guerreiro, V. Vedral, A. Zeilinger, and M. Aspelmeyer, Optomechanical Entanglement between a Movable Mirror and a Cavity Field, Phys. Rev. Lett. {\bf 98}, 030405 (2007).

\bibitem{Frohlich}H. Fr\"{o}hlich, Theory of the Superconducting State. I. The Ground State at the Absolute Zero of Temperature, Phys. Rev. {\bf 79}, 845 (1950).

\bibitem{Nakajima}S. Nakajima, Perturbation theory in statistical mechanics, Adv. Phys. {\bf 4}, 363 (1953).

{\bibitem{Aggarwal2020}N. Aggarwal, T.J. Cullen, J. Cripe, G. D. Cole, R. Lanza, A. Libson, D. Follman, P. Heu, T. Corbitt, and N. Mavalvala, Room-temperature optomechanical squeezing, Nat. Phys. {\bf 16}, 784 (2020).}

{\bibitem{Han2020}X. Han, W. Fu, C. Zhong, C. L. Zou, Y. Xu, A. A. Sayem, M. Xu, S. Wang, R. Cheng, L. Jiang, and H. X. Tang, Cavity piezo-mechanics for superconducting-nanophotonic quantum interface, Nat Commun {\bf 11}, 3237 (2020).}

{\bibitem{Zhao2022}H. Zhao, A. Bozkurt, and M. Mirhosseini, Electro-optic transduction in silicon via GHz-frequency nanomechanics, arXiv:2210.13549}

\bibitem{Boneberg-2022}M. Boneberg, I. Lesanovsky, and F. Carollo, Quantum fluctuations and correlations in open quantum Dicke models, Phys. Rev. A {\bf 106}, 012212 (2022).

\bibitem{Ballestero-2020}C. Gonzalez-Ballestero, J. Gieseler, and O. Romero-Isart, Quantum Acoustomechanics with a Micromagnet, Phys. Rev. Lett. {\bf 124}, 093602 (2020).

\bibitem{LiB-2019}B. Li, P. B. Li, Y. Zhou, J. Liu, H. R. Li, and F. L. Li, Interfacing a Topological Qubit with a Spin Qubit in a Hybrid Quantum System, Phys. Rev. Appl. {\bf 11}, 044026 (2019).

\bibitem{Angerer17} A. Angerer, S. Putz, D. O. Krimer, T. Astner, M. Zens, R. Glattauer, K. Streltsov, W. J. Munro, K. Nemoto, S. Rotter, J. Schmiedmayer, and J. Majer, Ultralong relaxation times in bistable hybrid quantum systems, Sci. Adv. {\bf 3}, e1701626 (2017).

{\bibitem{Johansson1} J. R. Johansson, P. D. Nation, and F. Nori, Qutip: An open-source Python framework for the dynamics of open quantum systems, Comput. Phys. Commun. {\bf 183}, 1760 (2012).}

{\bibitem{Johansson2} J. R. Johansson, P. D. Nation, and F. Nori, Qutip2: A Python framework for the dynamics of open quantum systems, Comput. Phys. Commun. {\bf 184}, 1234 (2013).}
\end{thebibliography}
\end{document}